\documentclass[aps,pra,reprint,twocolumn,superscriptaddress,showpacs,showkeys,superscriptaddress,longbibliography]{revtex4-1}
\usepackage{amsmath,amssymb,amstext}
\usepackage[usenames,dvipsnames]{color}
\usepackage{graphicx}
\usepackage{bm,bbold,braket}
\usepackage{natbib}
\usepackage{txfonts, comment,stmaryrd}
\usepackage{dcolumn}
\usepackage{color}
\usepackage{xcolor}
\colorlet{rn}{red}
\colorlet{an}{blue}
\usepackage[english]{babel}
\DeclareMathOperator{\Tr}{Tr}

\usepackage[colorlinks,bookmarks=false,citecolor=blue,linkcolor=red,urlcolor=blue]{hyperref}
\begin{document}

\title{Dynamics and Quantum correlations in Two independently driven Rydberg atoms with distinct laser fields}
\author{Vineesha Srivastava}
\affiliation{Indian Institute of Science Education and Research, Pune 411 008, India}    
\affiliation{Indian Institute of Technology (BHU), Varanasi 221 005, India} 
\author{Ankita Niranjan}
\affiliation{Indian Institute of Science Education and Research, Pune 411 008, India}    
\author{Rejish Nath}
\affiliation{Indian Institute of Science Education and Research, Pune 411 008, India}    

%\date{\today}

\begin{abstract}
We study the population dynamics in a two-atom setup in which each atom is  driven independently by different light fields, but coupling the same Rydberg state. In particular, we look at how an offset in the Rabi frequencies between two atoms influences the dynamics. We find novel features such as amplifying the Rabi frequency of one atom, together with strong Rydberg-Rydberg interactions freezes the dynamics in the second atom. We characterize the Rydberg-biased freezing phenomenon in detail, with effective Hamiltonians obtained for various limits of the system parameters. In the absence of Rabi-offset, the doubly excited state population exhibits a Lorentzian profile as a function of interaction, whereas for very small offsets it shows splitting and thus peaks. Using an effective Hamiltonian as well as the perturbation theory for weak interactions, we show that the peak arises from a competition between Rabi-offset and Rydberg-Rydberg interactions when both are sufficiently small, together with the Rydberg blockade at large interactions. The effective Hamiltonians provide us with analytical results which are in an excellent agreement with full numerical solutions. Also, we analyze the growth and the dynamics of quantum correlations such as entanglement entropy and quantum discord for the coherent dynamics.  We extend our studies to the dissipative case in which the spontaneous emission from the Rydberg state is taken into account and in particular, we look at the purity and quantum discord of the steady states. To conclude, our studies reveal that the local manipulation of an atom using Rabi-offset can be an ideal tool to control the quantum correlations and in general, quantum states of the composite two-qubit systems.
\end{abstract}



\maketitle
%\tableofcontents

\section{Introduction}
% \cite{urban2009observation} \cite{gaetan2009observation} \cite{wilk2010entanglement} 
Rydberg excited atoms have emerged as a great platform to test and study various quantum phenomena \cite{saf10,low12,bro16,jon17}, due to the multifaceted nature in engineering their properties using external fields. In particular, the strong Rydberg-Rydberg interactions \cite{beg13} lead to a well-known phenomenon called the Rydberg or dipole blockade \cite{luk01,gae09, urb09} and is a crucial mechanism for applications from quantum many-body physics \cite{wei10,sch12,bar15,sch15,zei16,ber17,zei17,mar17,gro17} to quantum information protocols \cite{jak00,saf10,wil10,ise10,saf17}. 

A typical Rydberg setup has been modeled as a gas of interacting two-level atoms (qubits), with interactions of either dipolar or van der Waals type. The minimal setup constitutes of either two atoms or two excitations and has been a usual scenario in many of the Rydberg-based experimental studies \cite{urb09,gae09, wil10,ise10,rya10,beg13,syl14,lab14,syl15,jau15,syl17,zen17,pic18,lev18}. On an equivalent note, the atom-based technologies have progressed in such a way that it is possible to probe and manipulate at a single particle level \cite{ott16, kuh16,gro17}. Recently, involving  Rydberg excitations, single atom addressing has been used to quantify the imperfections in Rabi oscillations \cite{del18}, proposed to engineer phonon modes in ion crystals \cite{wli13, Nat15}, to achieve controlled local operations to manipulate two atom quantum state \cite{lab14}, controlling resonant dipole-dipole interaction between Rydberg atoms \cite{bro16}, as well as freezing spin excitation dynamics \cite{syl17} and controlled quantum gates \cite{mal15,lev18}.

On the other side, to characterize quantum correlations in composite systems is a nontrivial and an important task for developing scalable quantum technologies. In particular, entanglement entropy and quantum discord have gained vital importance because of their applications in condensed matter physics, especially in characterizing various quantum phases \cite{eis10,mod12,ami08,dil08,lat09,che10,wan12,ani18}, including topologically ordered ones \cite{lev06,kit06,jia12,che10} and spin liquids \cite{isa11,zha11}. While entanglement entropy measures the entanglement between two subsystems \cite{ben96,benn96} in pure states, discord is treated as a more general non-classical correlation, may possess a nonzero value even for separable mixed states  \cite{hen01, oll01} and is highly relevant for deterministic quantum computation with a single qubit \cite{dat08,lan08, okr11, ani18}. Associated to entanglement entropy in pure states, an alternate measure for entanglement in mixed states is the entanglement of formation \cite{ben96}. So far, using ultracold atomic setups, the dynamics of entanglement (R\'enyi) entropy has been measured in quantum quench experiments \cite{raj15,kau16}.

\begin{figure}
\centering
\includegraphics[width= 1.\columnwidth]{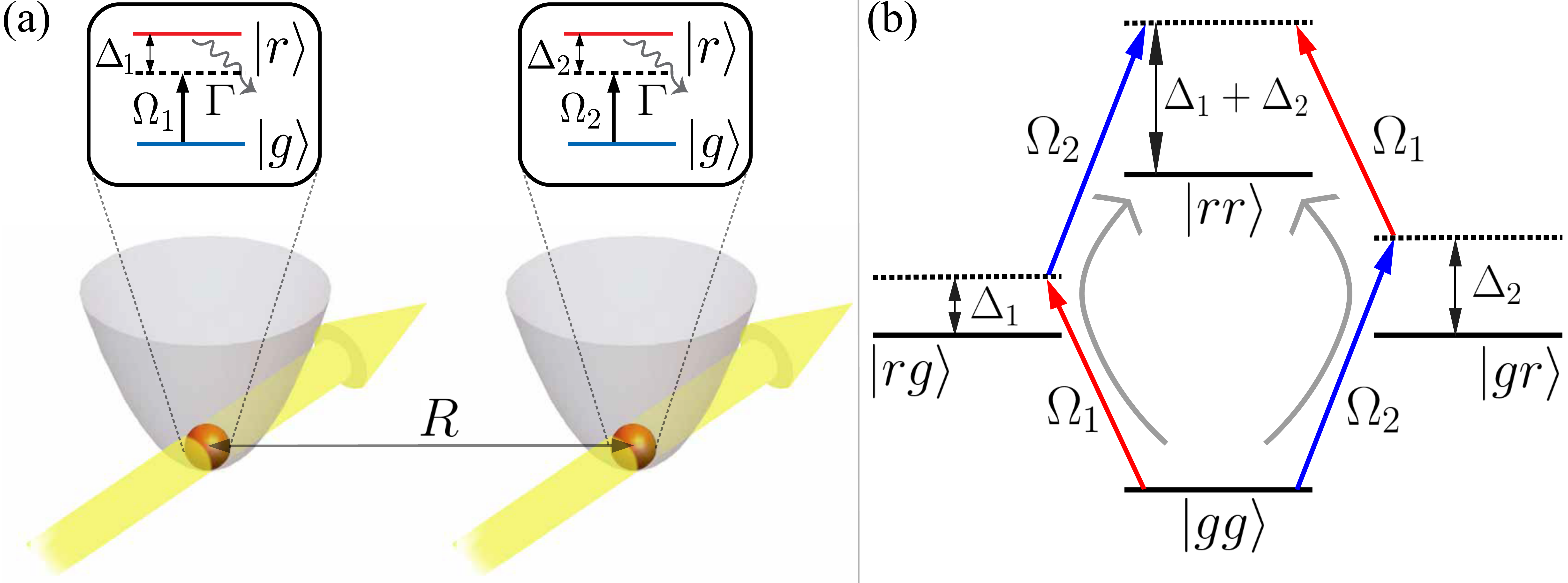}
\caption{(a) Schematic picture of two atoms trapped using microtraps with a separation of $R$. Each of the atoms is assumed to be driven independently by different laser fields coupling the ground state $|g\rangle$ to the same Rydberg state $|r\rangle$.  $\Gamma$ is the decay rate from the Rydberg state $|r\rangle$. (b) shows the level scheme for two atoms and the corresponding laser parameters associated with each state including the Rabi couplings. The two primary trajectories by which the state $|rr\rangle$ is populated from $|gg\rangle$, via $|gr\rangle$ and $|rg\rangle$ are shown by curved arrows.}
\label{fig1}
\end{figure}

In this paper, motivated by above developments on single particle control and manipulation, we study a minimal setup of two two-level Rydberg atoms driven continuously and independently by two distinct laser fields. In particular, we look at the effect of an offset in Rabi frequencies between the fields on the population dynamics, in the presence of Rydberg-Rydberg interactions. Interestingly, the population dynamics reveals us qualitatively novel features. An interesting scenario emerges when amplifying the Rabi coupling in one atom freezes the dynamics in the second atom, in the presence of strong atom-atom interactions. This phenomenon we term it as {\em Rydberg-biased freezing} and is also well captured by effective Hamiltonians at different limits of system parameters. Without the Rabi-offset, the time-averaged doubly excited state population exhibits a Lorentzian profile as a function of interaction strength. Whereas for small offset, the Lorentzian profile gets deformed, showing a non-monotonous behaviour with a peak at small interaction strengths. Anew, obtaining an effective Hamiltonian as well as using the second order perturbation theory, we show that the peak arises from a competition between Rabi-offset and Rydberg-Rydberg interaction when both are sufficiently small, together with Rydberg blockade at large interactions.  The striking quality of effective Hamiltonians obtained is that we can attain analytical solutions for population dynamics at various limits. Further, we analyze the growth and temporal evolution of quantum correlations such as the entanglement entropy and the quantum discord for the coherent dynamics. While entanglement entropy serves as a good measure for quantum correlations only in pure states,  quantum discord is used for both pure and mixed states. The correlation dynamics also reveals the competition between the inter-atomic interactions and Rabi-offset when both are sufficiently small. We extend our studies to the dissipative case in which the spontaneous emission from the Rydberg state is taken into account and in particular, we look at the purity and quantum discord in the steady states \cite{ahu13,fan17}. Finally, we conclude and demonstrate that the Rabi-offset can constitute an ideal tool to control quantum correlations between the two qubits.

The paper is structured as follows. In Sec. \ref{sup} we discuss the schematic setup, the model Hamiltonian, the master equation and define the quantum correlations that we analyze. In Sec. \ref{dyn1} we analyze the coherent dynamics of the system as a function of Rabi-offset and interaction strength. In Sec. \ref{heffs} we obtain effective Hamiltonians describing the features of the dynamics in various limits of system parameters. In Sec. \ref{corr} we study the dynamics of quantum correlations such as entanglement entropy and quantum discord for the coherent dynamics discussed in Sec. \ref{dyn1}. Finally, in Sec. \ref{diss} we consider the spontaneous emission from the Rydberg state and look at the dynamics and steady-state quantum correlations. In the appendix, we provide detailed calculations on the perturbation theory in the weak interactions limit as well as analytical results for the steady state density matrices and purity for the system and subsystems.

%%%%%%%%
\section{setup, Model and Quantum correlations}
%%%
\label{sup}
We consider two two-level atoms or qubits, each of them being strongly confined in two independent micro traps [see Fig. \ref{fig1}(a)], and are driven by distinct laser fields with Rabi frequencies $\Omega_i$ and detunings $\Delta_i$, coupling the electronic ground state $|g\rangle$ to a Rydberg state $|r\rangle$. Let $V_0=C_6/R^6$ provides us the strength of the van der Waals interaction between the  two Rydberg excited atoms. In the frozen gas limit \cite{mou98,and98}, ignoring the motional dynamics of the atoms, the internal state dynamics of the setup is governed by the Hamiltonian ($\hbar=1$):
\begin{equation}
\hat H=-\sum_{i=1}^2\Delta_i\hat\sigma_{rr}^{i}+\sum_{i=1}^2\frac{\Omega_i}{2}\hat\sigma_x^{i}+V_0\hat\sigma_{rr}^{1}\hat\sigma_{rr}^{2},
\label{ham}
\end{equation}
where $\hat\sigma_{ab}=|a\rangle\langle b|$ with $a, b\in \{r, g\}$, $\hat\sigma_x=\hat\sigma_{rg}+\hat\sigma_{gr}$. Introducing $\Delta_1=\Delta$, $\Delta_2=\Delta+\delta$, $\Omega_1=\Omega$ and $\Omega_2=\Omega+\omega$ with $\omega>0$, we rewrite the Hamiltonian in Eq. (\ref{ham}) as,
\begin{equation}
\hat H=-\Delta\sum_{i=1}^2\hat\sigma_{rr}^{i}+\frac{\Omega}{2}\sum_{i=1}^2\hat\sigma_x^{i}+V_0\hat\sigma_{rr}^{1}\hat\sigma_{rr}^{2}-\delta\hat\sigma_{rr}^2+\frac{\omega}{2}\hat\sigma_x^2.
\label{ham2}
\end{equation}
 In the absence of Rabi couplings ($\Omega=\omega=0$), the eigenstates of the Hamiltonian in Eq. (\ref{ham2}) are $|gg\rangle$, $|rg\rangle$, $|gr\rangle$ and $|rr\rangle$ with eigenvalues $E_{gg}=0$, $E_{rg}=-\Delta$, $E_{gr}=-\Delta-\delta$ and $E_{rr}=-2\Delta-\delta+V_0$. The coherent dynamics of the system is obtained by numerically solving the Schr\"odinger equation: $i\partial/\partial t |\psi(t)\rangle=\hat H|\psi(t)\rangle$ and we use the two-atom basis  $\{|gg\rangle, |gr\rangle, |rg\rangle, |rr\rangle\}$. Throughout we take $|\psi(t=0)\rangle=|gg\rangle$ and assume both the fields are at resonance ($\delta=\Delta=0$) with $|g\rangle$ - $|r\rangle$ transition and we focus on the effect of Rabi-offset $\omega$ on the population dynamics. With above assumption, the Hamiltonian reduces to 
 \begin{equation}
 \hat H=\frac{\Omega}{2}\sum_{i=1}^2\hat\sigma_x^{i}+\frac{\omega}{2}\hat\sigma_x^2+V_0\hat\sigma_{rr}^{1}\hat\sigma_{rr}^{2}.
 \label{ham3}
 \end{equation}
 The level scheme of our two atom setup with laser parameters is shown in Fig. \ref{fig1}(b) and the two primary trajectories by which the state $|rr\rangle$ is populated from $|gg\rangle$, via $|gr\rangle$ and $|rg\rangle$ are shown by curved arrows. When $\omega$ vanishes, the system is identical to the scenario in which both the atoms are driven by a global field. In that case, for sufficiently large $V_0 (>\Omega)$ starting from $|gg\rangle$, the system undergoes coherent Rabi oscillations between $|gg\rangle$ and $|+\rangle=(|gr\rangle+|rg\rangle)/\sqrt{2}$ with an enhanced Rabi frequency of $\sqrt{2}\Omega$. These dynamics is attributed to the Rydberg blockade. A non zero $\omega$ breaks the symmetry between the two atoms and hence, the state $|+\rangle$ lost its significance in the dynamics. 

Apart from the state dynamics, we also look at how the quantum correlations develop and evolve in our setup in particular, entanglement entropy and quantum discord. Since the initial state $|gg\rangle$ is not an eigenstate of the Hamiltonian in Eq. (\ref{ham3}), our scenario is identical to that of a quantum quench problem in which the Rabi frequencies are instantaneously quenched from zero to a finite value at $t=0$.  Let us label the first atom as subsystem $A$ and the strongly driven second atom as subsystem $B$. The entanglement entropy of the subsystems are, $\mathcal S_A=-\Tr(\hat\rho_A\log_2 \hat\rho_A)$ and $\mathcal S_B=-\Tr(\hat\rho_B\log_2 \hat\rho_B)$ where $\hat\rho_A$ ($\hat\rho_B$) is the reduced density matrix for $A$ ($B$), which is obtained by the partial trace of the total density matrix, $\hat \rho=|\psi(t)\rangle\langle \psi(t)|$. The partial trace over $\rho$ leaves the subsystems $A$ and $B$ in a classical mixture of pure quantum states. In terms of the eigenvalues ($\lambda_i$) of $\hat\rho_A$, we have $\mathcal S_A=-\sum_{i=1}^2\lambda_i\log_2\lambda_i$, and for a pure state $\mathcal S_A=\mathcal S_B$. 

To define the quantum discord, we briefly sketch the mutual information in the classical information theory. The classical mutual information between two subsystems $A$ and $B$ is defined as $\mathcal I= H_A+H_B-H_{AB}$ where $H_A$ ($H_B$) is the Shannon entropy of the subsystem $A$ ($B$), and $H_{AB}$ is the joint entropy of $A$ and $B$. An equivalent expression for mutual information is $\mathcal J=H_B-H_{B|A}$, where $H_{B|A}$ is the conditional entropy, the information needed to describe $B$ when $A$ is known. Though, in the classical theory $\mathcal I=\mathcal J$, in the quantum version, in which the Shannon entropy is replaced by the von Neumann entropy, there exists a discrepancy between $\mathcal I$ and $\mathcal J$ which is quantified by the quantum discord.

\begin{figure}
\vspace{0.cm}
\centering
\includegraphics[width= 1.\columnwidth]{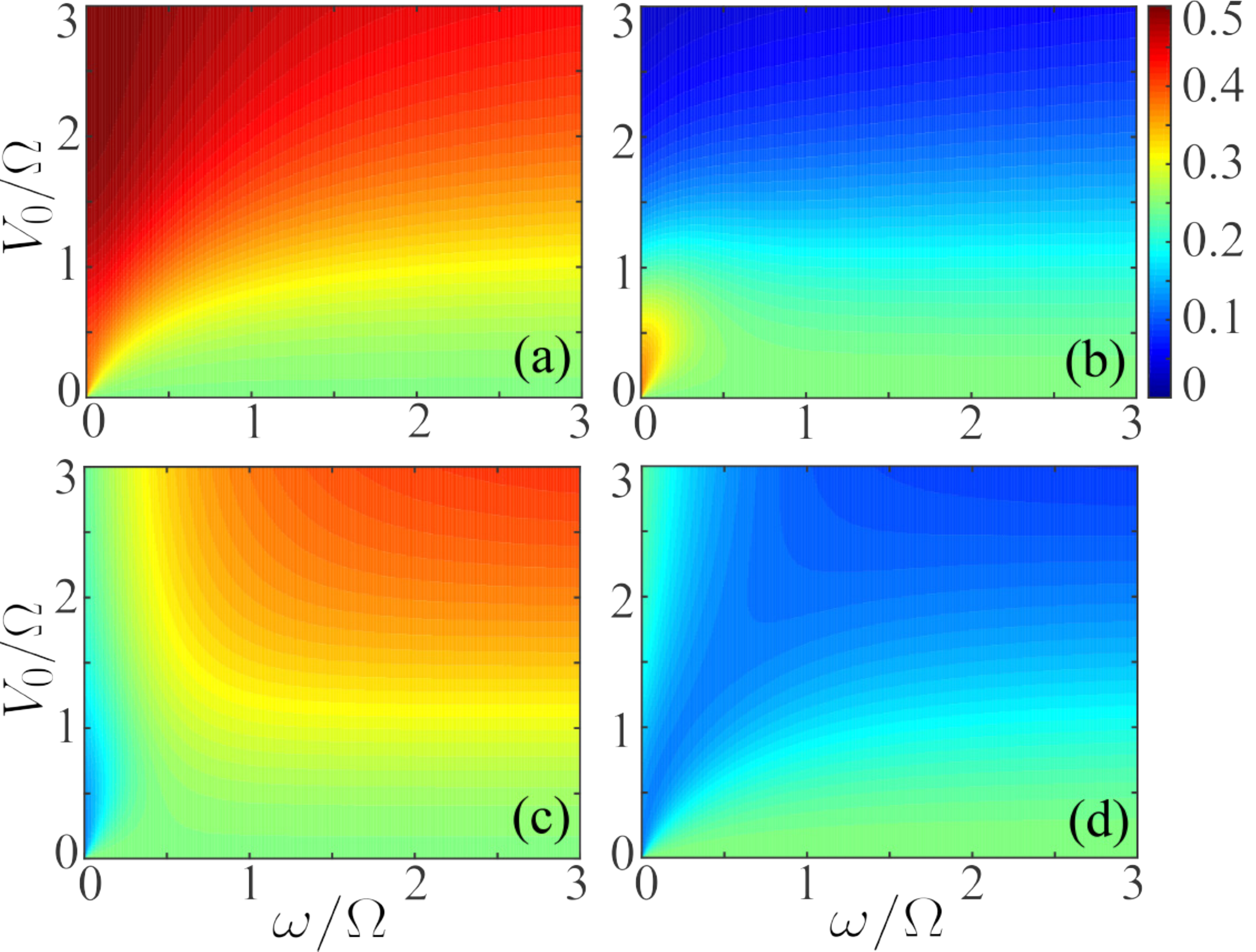}
\caption{\small{(Color online) Density plot for time average populations: (a) $\bar P_{gg}$, (b) $\bar P_{rr}$, (c) $\bar P_{gr}$ and (d) $\bar P_{rg}$ as a function of $V_0/\Omega$ and $\omega/\Omega$ for $\Delta = \delta = 0$ and $|\psi(t=0)\rangle=|gg\rangle$ at $\Omega\tau=5000$. For large values of $V_0/\Omega$ and $\omega/\Omega$ both $\bar P_{rr}$ and $\bar P_{rg}$ are negligible, leading to the phenomenon of Rydberg-biased freezing. Also (b) reveals us that the blockade condition is weakly affected by the Rabi-offset $\omega$.}}
\label{fig:2} 
\end{figure}

 In the quantum theory, we have $\mathcal I= S_A+S_B-S_{AB}$ and $\mathcal J(B:A)=S_B-S_{B|A}$, where $\mathcal S_{AB}=-\Tr(\hat\rho\log_2\hat\rho)$ is the von Neumann entropy for the state $\hat\rho$, and $\mathcal S_{AB}=0$ for a pure state. Given a complete set of von Neuman projective measurements $\{\hat \Pi_A^i\}$ on the subsystem $A$ with probabilities $\{p^i\}$, the conditional entropy of the subsystem $B$ is $\mathcal S_{B|A}=\sum_ip^i\mathcal S_{B|i}$, where $\mathcal S_{B|i}$ is the von Neumann entropy for the reduced density operator $\hat\rho_{B}^i= \Tr_{A} \left[(\hat\Pi_{A}^i\otimes\mathbb I_B)\hat\rho_{AB}(\hat\Pi_{A}^i\otimes\mathbb I_B)^{\dagger}\right]/p^i$ with $p^i= \Tr_{AB} \left[(\hat\Pi_{A}^i\otimes\mathbb I_B)\hat\rho_{AB}(\hat\Pi_{A}^i\otimes\mathbb I_B)^{\dagger}\right]$ and $\mathbb I_B$ is the identity operator. It has been shown that the total classical correlation can be obtained as, $\tilde{\mathcal J}(B:A)=\max_{\{\hat\Pi_A^i\}}\left[S_B-\sum_ip^i\mathcal S_{B|i}\right]$ \cite{hen01}. The maximization ($\max_{\{\hat\Pi_A^i\}}$) is carried across all the possible orthonormal measurement bases $\{\hat\Pi_A^i\}$ of the subsystem $A$. Similarly one can obtain $\tilde{\mathcal J}(A:B)$  where the measurements are being carried out on the subsystem $B$. Finally, the quantum discord is defined on both ways by swapping $A$ and $B$ as, 
\begin{equation}
 \mathcal D(A:B)=\mathcal I-\tilde{\mathcal J}(A:B), 
 \end{equation}
 and 
 \begin{equation}
 \mathcal D(B:A)=\mathcal I-\tilde{\mathcal J}(B:A).
 \end{equation}
 Note that, the quantum conditional entropy depends on the choice of the observables being measured on the other subsystem, and this results in a discrepancy between $\mathcal I$ and $\mathcal J(B:A)$ or $\mathcal J(A:B)$, which is quantified as the quantum discord. For a bipartite pure state $|\psi(t)\rangle$ the quantum discord coincides with the entanglement entropy, i.e.,  $\mathcal D(A:B)=\mathcal D(B:A)=\mathcal S_A=\mathcal S_B$ \cite{ade10}.

Once the spontaneous emission from the Rydberg state is taken into account, the dissipative dynamics and the steady state correlations are analyzed using the master equation for the two-particle density matrix $\hat \rho$,
\begin{equation}
\partial_t \hat{\rho} = -i \left[\hat{H},\hat{\rho}\right]  +\mathcal{L} [\hat{\rho}],
\label{meq}
\end{equation}
with the Lindblad operator given by 
\begin{equation}
\mathcal{L}[\hat \rho] =   \sum_{i=1}^2\hat C_i \hat{\rho} \hat C^{\dagger}_i- \frac{1}{2} \sum_i \left(\hat C^{\dagger}_i \hat C_i \hat{\rho} + \hat{\rho}\hat C^{\dagger}_i \hat C_i\right)
\end{equation}
where the operator, $\hat C_i= \sqrt{\Gamma} \hat\sigma_{ge}^i$ with $\Gamma$ as the spontaneous decay rate of the Rydberg state $|r\rangle$.  At the steady state, $\partial_t \hat{\rho} =  0$ and the dissipative mechanism drives the system eventually into a mixed state even though the system is initially prepared in a pure state. Note that, for a mixed state the discord is no longer the same as the entanglement entropy and infact, $\mathcal S_{A,B}$  has been ruled out from being a good measure of quantum correlations or entanglement since it fails to distinguish between classical and quantum correlations  \cite{ben96, ved97, ved98}, whereas  the quantum discord remains a good measure. For a mixed state in general, $\mathcal D(A:B)\neq \mathcal D(B:A)$ since the conditional entropy is not symmetric for all states \cite{mod12}. The exception is, i.e. $\mathcal D(A:B)=\mathcal D(B:A)$ when the states are symmetric under the exchange of $A$ and $B$.

%%%

\section{State Population Dynamics}
\label{dyn1}
%%%

\begin{figure}
\centering
\includegraphics[width=1.\columnwidth]{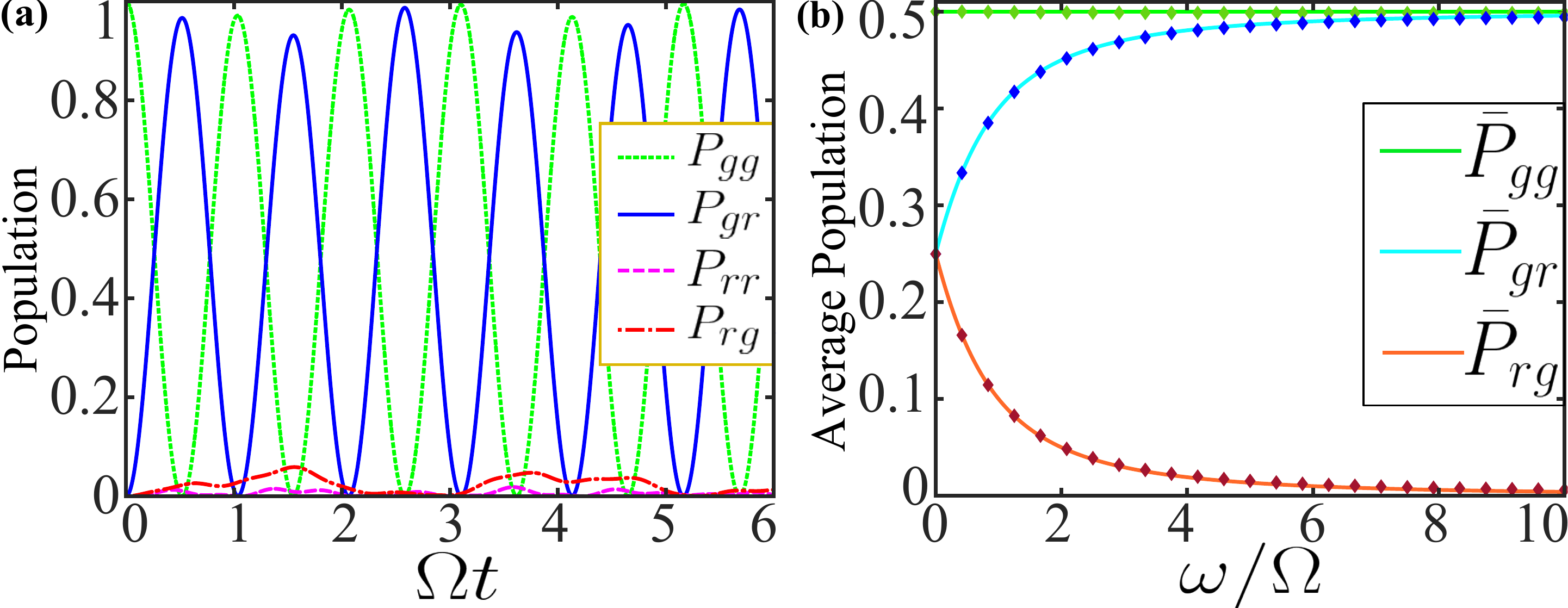}
\caption{(a) The population dynamics at $V_0/\Omega=10$ and $\omega/\Omega=5$. (b) The time average populations as a function of $\omega/\Omega$ for $V_0=25\Omega$, $\delta=0$ and $\Omega\tau=5000$. Since $V_0\gg \Omega$, the numerical results (solid points) are in excellent agreement with the analytic results (solid lines) given in Eqs. (\ref{p1})-(\ref{p2}).}
\label{fig:3}
\end{figure}

Here, we analyze the effect of Rabi-offset $\omega$ on the Rydberg excitation dynamics, in particular, how it affects the Rydberg blockade. We look at the time averaged populations (see Figs. \ref{fig:2} and \ref{fig:3}): $\bar P_{\alpha\beta}=1/\tau\int_0^\tau P_{\alpha\beta}(t) dt$ with $\alpha, \beta \in \{r, g\}$ of the states $\{|\alpha\beta\rangle\}$, as a function of $\omega$ and $V_0$. For $\omega=0$ and sufficiently large $V_0/\Omega$, the interaction induced level shift in $|rr\rangle$ state results in the well known Rydberg blockade with average populations approaching $\bar P_{gg}\to 0.5$, $\bar P_{rr}\to 0$, $\bar P_{+}\to 0.5$, where $P_{+}=P_{rg}+P_{gr}$. For $\omega\ll\Omega$, we have $\bar P_{gr}\approx \bar P_{rg}$ independently of $V_0$. When $\omega$ is significantly large, the results shown in Fig. \ref{fig:2} reveal interesting features. The first thing to notice from Fig. \ref{fig:2}(b) is that the blockade condition is merely affected by $\omega$ for sufficiently large $\omega$. This is understood as follows: as $\omega$ becomes large, the second atom is driven strongly compared to the first one, and that results in the augmentation of $P_{gr}$ at the cost of $P_{rg}$. Hence, to attain blockade the Rydberg-Rydberg interactions just have to dominate the Rabi coupling of the weakly driven atom, i.e., $V_0>\Omega$, leaving the blockade condition almost independent of $\omega$, which become more apparent in Sec. \ref{heffs}. It also implies that, the blockade dynamics not necessarily always result in the generation of the symmetric entangled state $|+\rangle$ as in the case for $\omega=0$. Summing up, we have an interesting scenario: for sufficiently large $\omega/\Omega$ and $V_0/\Omega$, the dynamics of the first atom nearly freezes and the system exhibits coherent Rabi oscillations between $|gg\rangle$ and $|gr\rangle$ with $P_{rr} \approx 0$ and $P_{rg}\approx 0$ as shown in Figs. \ref{fig:3}. Note that, the freezing of the first atom emerges as a combined effect of both the Rydberg blockade from large interactions and the strong driving in the second atom, and we term this as {\em Rydberg-biased freezing}.

%%%%%%
%%%%%
%%%%%

\begin{figure}
\vspace{0.cm}
\centering
\includegraphics[width= 1.\columnwidth]{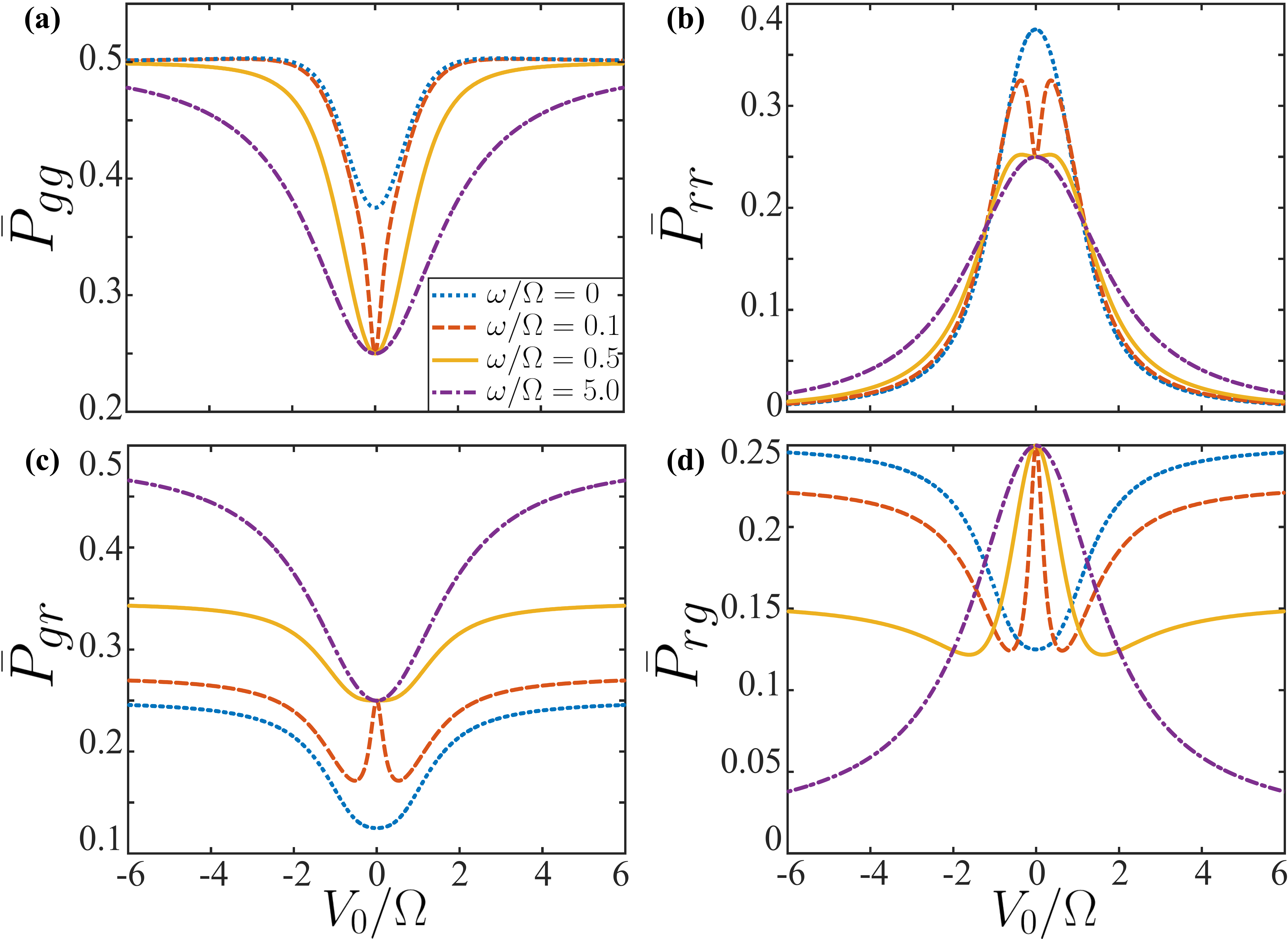}
\caption{\small{(Color online)The time average populations: (a) $\bar P_{gg}$, (b) $\bar P_{rr}$, (c) $\bar P_{gr}$ and (d) $\bar P_{rg}$ as a function of $V_0/\Omega$ for different $\omega/\Omega$, with $\Delta = \delta = 0$, $|\psi(t=0)\rangle=|gg\rangle$ and $\Omega\tau=5000$. For very small values of  $\omega/\Omega$, the populations $\bar P_{rr}$, $\bar P_{rg}$ and $\bar P_{gr}$ depend non-monotonously on $|V_0|$.}}
\label{fig:4} 
\end{figure}

To gain more insights into the population dynamics, in Fig. \ref{fig:4} we show explicitly the average populations as a function of $V_0/\Omega$ for different values of $\omega/\Omega$. When $V_0=0$, the two atom states are simply the product of single atom states and we have $P_{gg}(t)=\cos^2\frac{\Omega t}{2}\cos^2\frac{(\Omega+\omega)t}{2}$, $P_{gr}(t)=\cos^2\frac{\Omega t}{2}\sin^2\frac{(\Omega+\omega)t}{2}$, $P_{rg}(t)=\sin^2\frac{\Omega t}{2}\cos^2\frac{(\Omega+\omega)t}{2}$ and $P_{rr}(t)=\sin^2\frac{\Omega t}{2}\sin^2\frac{(\Omega+\omega)t}{2}$ for the initial state $|\psi(t=0)\rangle=|gg\rangle$. The corresponding time average values are $\bar P_{gg}=\bar P_{rr}=0.375$,  $\bar P_{gr}=\bar P_{rg}=0.125$ for $\omega=0$ and $\bar P_{gg}=\bar P_{rr}=\bar P_{gr}=\bar P_{rg}=0.25$ for $\omega\neq 0$. The effect of a nonzero $\omega$ on the time average populations at $V_0=0$ has an influence as well as when $V_0\neq 0$, and qualitatively noticeable especially, at small $\omega$ and $V_0$ as seen in Fig. \ref{fig:4}. 

%%%
%%%%%%

\begin{figure}
\centering
\includegraphics[width= 1.\columnwidth]{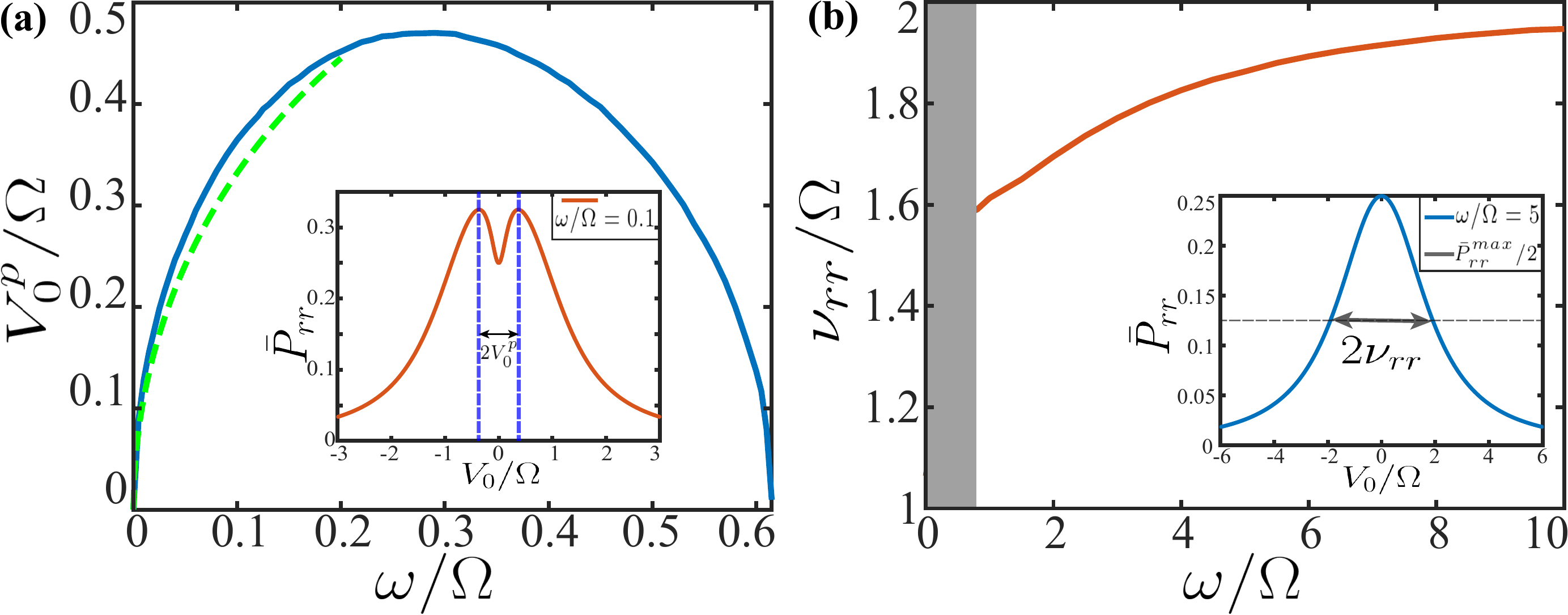}
\caption{(a) The half of the peak separation  $V_0^P$ in $\bar P_{rr}$ (see inset) as a function of $\omega/\Omega$. The dashed line is the analytical result given by Eq. (\ref{vop}) obtained for small $\omega$. (b) shows the Lorentzian width ($\nu_{rr}$) of $\bar P_{rr}$ as a function of $\omega/\Omega$. The width is obtained at the half-maximum of $\bar P_{rr}$ distribution in $V_0/\Omega$. The shaded region for small $\omega$ indicates the existence of double peak structure in $\bar P_{rr}$ shown in the inset of (a).}
\label{fig:5}
\end{figure}

As previously mentioned, when $\omega=0$ and $V_0\neq 0$ , the population is transferred from $|gg\rangle$ to $|rr\rangle$ state via the entangled state $|+\rangle$, thus preserving the exchange symmetry between the two atoms at any instant, and $\bar P_{rr}$ decreases monotonously with increasing $|V_0|$ whereas $\bar P_{gg}$ and $\bar P_{+}$ increases and saturates to 0.5 at large $|V_0|$. As soon as $\omega\neq 0$, the symmetry is broken, resulting in $\bar P_{gr}\neq\bar P_{rg}$ and as expected, $\bar P_{gr}$ gets larger with larger $\omega$ for any $V_0\neq 0$. As seen in Fig. \ref{fig:4}(b) [also in Fig. \ref{fig:2}(b)], for small values of $\omega$, $\bar {P}_{rr}$ exhibits a  non-monotonous behaviour as a function of $|V_0|$. The single Lorentzian profile centered at $V_0=0$ of $\bar P_{rr}$ for $\omega=0$ exhibits a partial splitting, exhibiting two peaks at $\pm V_0^p$ around $V_0=0$. This arises from the competition between the terms associated with $V_0$ and $\omega$ in Eq. (\ref{ham3}) for small values of $V_0$ and $\omega$. This is also evident from the plots of $\bar {P}_{gr}$, $\bar {P}_{rg}$, and $\bar {P}_{gg}$ shown in Fig. \ref{fig:4}. Focusing on $\bar {P}_{rr}$ [Fig. \ref{fig:2}(b)], for very small $\omega$ such that $\bar P_{gr}\approx\bar P_{rg}$, increasing $|V_0|$ from zero suppresses the effect of the offset $\omega$, leads to the recovery of  $\bar {P}_{rr}$ to the value obtained for $\omega=0$. Once $|V_0|$ dominates $\bar {P}_{rr}$ starts to decrease as expected from the blockade effect. As a result, the peaks separation $\left(2V_0^p\right)$ increases with increasing $\omega$ until it reaches a value for $\omega$ such that $V_0$ can no longer nullify the effect of $\omega$, see Fig. \ref{fig:5}(a). At that point, there is a substantial difference between the magnitudes of $\bar P_{gr}$ and  $\bar P_{rg}$, and $V_0^p$ decreases with further increase in $\omega$, reaches zero. Thus, when $\omega$ becomes greater than a particular value, the population $\bar {P}_{rr}$ becomes a Lorentzian function of $V_0$. The peak separation, $V_0^p\propto \sqrt{\omega}$ for small values of $\omega$. The width of the Lorentzian $\nu_{rr}$ as a function of $\omega$ is shown in Fig. \ref{fig:5}(b). The shaded region for small $\omega$ in Fig. \ref{fig:5}(b) indicates the existence of double peak structure, and $\nu_{rr}$ saturates to a constant value at large values of $\omega$. The latter also indicates that the blockade condition is not affected by $\omega$ at large values of $\omega$, as we have discussed above. In Sec. \ref{heffs}, using an effective Hamiltonian obtained in the limit $\{\omega, V_0\}\ll\Omega$, we explain the initial increment of $\bar {P}_{rr}$ in $V_0$ at small $\omega$ and also the same is done in Appendix \ref{a1} using the second order perturbation theory obtained in the weak interaction limit. Alternatively, one could also think, the peaks in $\bar P_{rr}$ vs $V_0$ emerge as a consequence of quantum interference since both $\omega$ and $V_0$ introduce additional phase shifts in the amplitudes of quantum paths populating the state $|rr\rangle$. Note that, the emergent splitting of $\bar {P}_{rr}$ in $V_0$ axis has a resemblance to the Autler-Townes effect, but here using two particle states and also  consequently the antiblockade effect \cite{ate07}. 
%%%%%%
%%%%%

\section{Effective Hamiltonians}
\label{heffs}
At this point, we obtain the effective Hamiltonians describing the long-time behaviour of our setup in different limits. First we consider the limit of $V_0\gg\{\Omega, \omega\}$ and we introduce the unitary transformation, $\hat U=\exp(iV_0\hat\sigma_{rr}^{1}\hat\sigma_{rr}^{2}t)$ to the Hamiltonian in Eq. (\ref{ham3}). The new Hamiltonian, $\hat H'=\hat U\hat H\hat U^{\dagger}+i(d\hat U/dt)\hat U^{\dagger}$ is,
\begin{align}
\begin{split}
        \hat{H'}= &\frac{\Omega}{2}\left[\ket{gg}\bra{gr} + \ket{gg}\bra{rg}+ e^{-i V_0 t}(\ket{gr}\bra{rr} + \ket{rg}\bra{rr})+ \rm{H.c.}\right]\\
        +&\frac{\omega}{2}\left(\ket{gg}\bra{gr}+ \ket{rg}\bra{rr}e^{-i V_0 t}+ \rm{H.c.}\right)\\
\end{split}
\end{align}
 In the second step, we obtain a period average Hamiltonian, $\hat H_{eff}^{(V_0)}=(1/T)\int_0^T H'(t)dt$, with time period $T=2\pi/V_0$, which provides us
\begin{equation}
\hat H_{eff}^{(V_0)}= \frac{\left(\omega+\Omega\right)}{2}\hat \sigma_{gg}^1\hat\sigma_x^2+\frac{\Omega}{2}\hat\sigma_x^1\hat\sigma_{gg}^2.
\label{hef}
\end{equation}
The second step is identical to removing the fast oscillations emerging from large $V_0$. Even after integrating out the interaction dependent terms, still the effective Hamiltonian $\hat H_{eff}^{(V_0)}$ cannot be written as a sum of two single particle terms, indicating the existence of quantum correlations between the two atoms, which we quantify later. The first term in Eq. (\ref{hef}) leaves the first atom in the ground state while driving the second atom with a Rabi coupling $\omega+\Omega$, and is vice versa for the second term but with a Rabi coupling $\Omega$ for the first atom. The two terms being the correlated Rabi couplings \cite{bas18}, also identical to the density assisted interband tunneling for atoms in optical lattices \cite{yas16}. When $\omega\gg\Omega$, the second term in Eq. (\ref{hef}) is merely a perturbation to the first term, leading to the scenario of Rydberg-biased freezing. Truncating the basis to $\{|gg\rangle, |gr\rangle, |rg\rangle\}$ and using the effective Hamiltonian in the Schr\"odinger equation, we can explicitly obtain the time dependent populations as,
\begin{eqnarray}
\label{p1}
P_{gg}(t)=\cos^2{\beta t} \\
\label{pm}
P_{gr}(t)=\left(\frac{\Omega+\omega}{2\beta}\right)^2\sin^2{\beta t} \\
P_{rg}(t)=\left(\frac{\Omega}{2\beta}\right)^2\sin^2{\beta t},
\label{p2}
\end{eqnarray}
where $\beta^2=\left[(\Omega+\omega)^2+\Omega^2\right]/4$. Note that, at $t=0$, $\hat U(0)=\mathcal I$, the identity operator which leaves the initial state unchanged, and is also same for the different transformations we consider below.  The time average populations become: $\bar P_{gg}=1/2$, $\bar P_{gr}=\left[(\Omega+\omega)/2\beta\right]^2/2$ and $\bar P_{rg}=(\Omega/2\beta)^2/2$. From Eqs. (\ref{pm}) and (\ref{p2}) we can see that for $\omega\ll\Omega$, we have $\bar P_{gr}\approx \bar P_{rg}$, which is consistent with numerical results shown in Fig. \ref{fig:2}. For $\omega\gg\Omega$ we have $P_{gr}(t)\approx\sin^2\beta t$ with $\beta\approx\omega/2$, indicating the Rabi oscillations between the states $|gg\rangle$ and $|gr\rangle$ with a Rabi frequency approximately $\omega$. We compare these results with the numerical solutions obtained by solving the full Hamiltonian in Eq. (\ref{ham3}), and is found to be in an excellent agreement when $V_0\gg\Omega$ [see Fig. \ref{fig:3}(b)].

%%%%%%
Now we derive the effective Hamiltonian in the limit  $\omega\gg \{\Omega, V_0\}$ by doing a similar procedure as above. For that we introduce a local unitary operator $\hat U=\exp(i\omega\hat\sigma_x^{i=2}t/2)$ acting only on the second atom which gives us the new Hamiltonian,
\begin{align}
\begin{split}
       \hat H'=&\frac{\Omega}{2}\sum_{i=1}^2\hat\sigma_x^{i} + V_0\left(\frac{1-\cos\omega t}{2}\hat\sigma_{rr}^1\hat\sigma_{gg}^2 + \frac{\cos\omega t + 1}{2}\hat\sigma_{rr}^1\hat\sigma_{rr}^2 \right. \\ \nonumber
        & \left. - \frac{\sin\omega t}{2}\hat\sigma_{rr}^1\hat\sigma_{y}^2\right),
\end{split}
\end{align}
and then averaging over a time period of $T=2\pi/\omega$, we get
\begin{equation}
\hat H_{eff}^{(\omega)}= \frac{\Omega}{2}\sum_{i=1}^2\hat\sigma_x^i+\frac{V_0}{2}\hat\sigma_{rr}^1.
\label{heff}
\end{equation}
The effective Hamiltonian $\hat H_{eff}^{(\omega)}$ can be written as a sum of two single particle Hamiltonians, with an effective detuning $V_0/2$ for the first atom. Note that, in the new rotating frame, the Rabi coupling of the second atom is reduced to $\Omega$. From $\hat H_{eff}^{(\omega)}$ it is clear that, for $V_0\gg\Omega$, the first atom is merely excited resulting in the Rydberg-biased freezing and also indicates that the two atom correlations get suppressed in the large $\omega$ limit (see Sec. \ref{corr}).

%%%%%%
\begin{figure}[hbt]
\centering
\includegraphics[width= 1.\columnwidth]{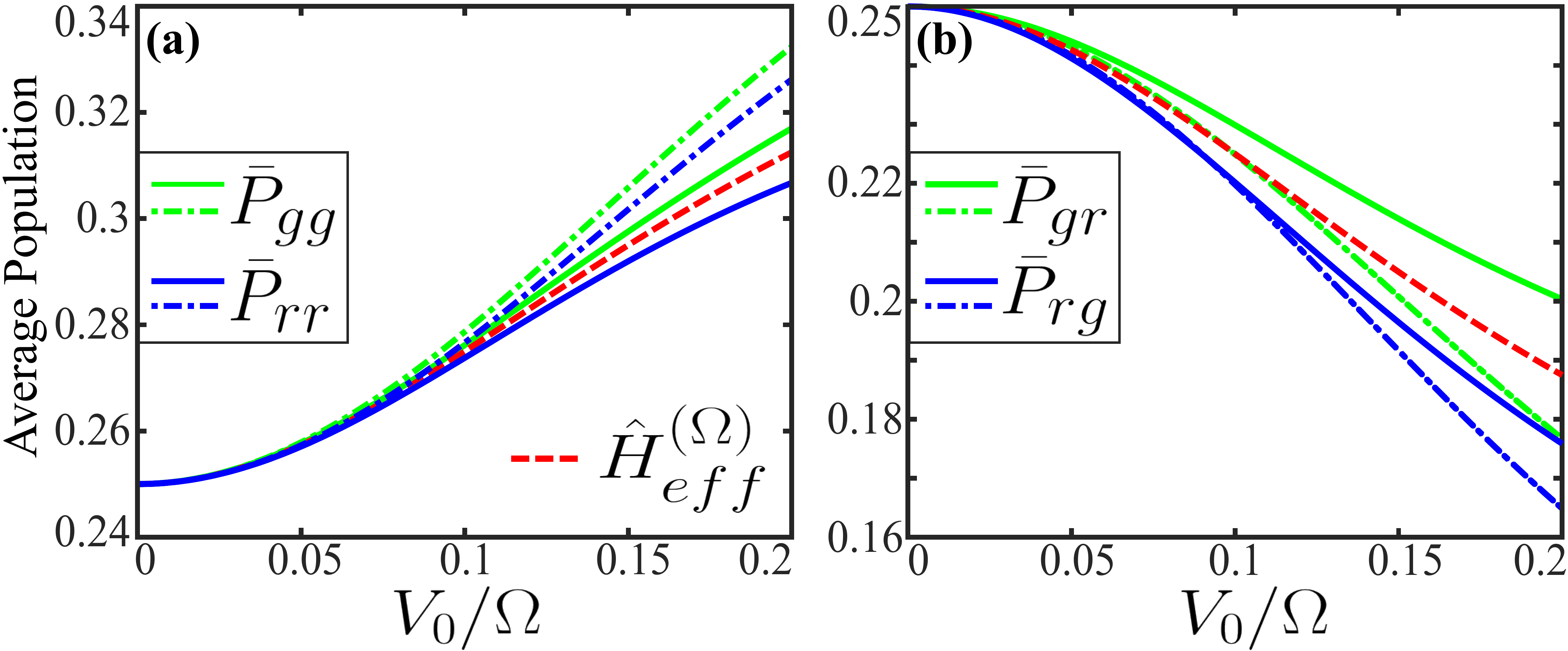}
\caption{Comparison of results from three different methods: exact numerical solution (solid lines), perturbation theory (dotted-dashed lines) for weak interactions and using the effective Hamiltonian $\hat H_{eff}^{(\Omega)}$ (dashed line) for $\omega/\Omega=0.1$. (a) is for the time average populations $\bar P_{gg}$ and $\bar P_{rr}$ and (b) is for $\bar P_{gr}$ and $\bar P_{rg}$.}
\label{fig:6}
\end{figure}
%%%%%

Finally, we consider the limit $\{V_0, \omega\}\ll \Omega$, and introduing the unitary operator $\hat U= e^{i\frac{\Omega}{2}(\hat\sigma_x^1+\hat\sigma_x^2) t}$ we obtain,
\begin{widetext}
\begin{eqnarray}
\nonumber
    \hat H'=& \frac{V_0}{4}\left[\left(\cos\Omega t-1\right)^2|gg\rangle\langle gg| + \left(\cos\Omega t+1\right)^2 |rr\rangle \langle rr| +\sin^2\Omega t\left((|gr\rangle \langle gr| + |rg\rangle \langle rg|)+ (|gr\rangle\langle rg| + \rm{H.c.})\right)+(\cos^2\Omega t- 1)(|gg\rangle\langle rr|+ \rm{H.c.})\right] \\ 
    & + \frac{\omega}{2}\hat\sigma_x^2+\left[\frac{iV_0\sin\Omega t}{4}\left(1-\cos\Omega t\right)(|gg\rangle\langle gr| + |gg\rangle\langle rg|)+ \rm{H.c.}\right]+ \left[\frac{iV_0\sin\Omega t}{4}\left(1+\cos\Omega t\right)(|gr\rangle\langle rr| + |rg\rangle\langle rr|)+ \rm{H.c.}\right].
    \end{eqnarray}
\end{widetext}
Then, the effective Hamiltonian after averaging over $T=2\pi/\Omega$, in the basis $\{|gg\rangle, |gr\rangle, |rg\rangle, |rr\rangle\}$:
\begin{equation}
    \hat H_{eff}^{(\Omega)}= 
    \begin{bmatrix}
    3V_0/8& \omega/2& 0& -V_0/8\\
    \omega/2& V_0/8& V_0/8& 0\\
    0& V_0/8& V_0/8& \omega/2\\
    -V_0/8& 0& \omega/2& 3V_0/8
    \end{bmatrix},
     \label{heff3}
\end{equation}
and together with the basis vectors transformations:
\begin{align}
 \nonumber   \hat U |gg\rangle =&\cos^2\frac{\Omega t}{2}|gg\rangle +\frac{i\sin\Omega t}{2}\left(|gr\rangle+|rg\rangle\right)-\sin^2\frac{\Omega t}{2}|rr\rangle\\ \nonumber
    \hat U |gr\rangle=&\frac{i\sin\Omega t}{2}\left(|gg\rangle+|rr\rangle\right) + \cos^2\frac{\Omega t}{2}|gr\rangle -\sin^2\frac{\Omega t}{2}|rg\rangle\\\nonumber
    \hat U |rg\rangle =&\frac{i\sin\Omega t}{2}\left(|gg\rangle+|rr\rangle\right) -\sin^2\frac{\Omega t}{2}|gr\rangle+ \cos^2\frac{\Omega t}{2}|rg\rangle\\
    \hat U |rr\rangle=&-\sin^2\frac{\Omega t}{2}|gg\rangle +\frac{i\sin\Omega t}{2}\left(|gr\rangle+|rg\rangle\right)+\cos^2\frac{\Omega t}{2}|rr\rangle,
    \label{rbs}
\end{align}
we can estimate the time average populations at small $\omega$ and $V_0$. For instance, when $V_0=\omega=0$ and with the initial state in the rotating frame, $|\psi_R(t=0)\rangle=\hat U|\psi(t=0)\rangle=|gg\rangle$, the time average populations can be estimated directly from $\hat U |gg\rangle$, which gives us $\bar P_{gg}=\bar P_{rr}=0.375$ and $\bar P_{gr}=\bar P_{rg}=0.125$, as expected. But, when either $\omega\neq 0$ or $V_0\neq 0$, we need to first use the Hamiltonian evolution of $\hat H_{eff}^{(\Omega)}$. Taking $V_0=0$ and $\omega\neq 0$, in the rotating frame, the initial $|gg\rangle$ state undergoes coherent Rabi oscillations with $|gr\rangle$ according to  $\hat H_{eff}^{(\Omega)}$, i.e., $|\psi_R(t)\rangle=\cos\frac{\omega t}{2}|gg\rangle-i\sin\frac{\omega t}{2}|gr\rangle$. Then, the population in $|rr\rangle$ state is obtained by projecting $|\psi_R(t)\rangle$ along $\hat U|rr\rangle$.  The latter provides us the same result discussed in Sec. \ref{dyn1}. Now, we take $\omega=0$ and $V_0\neq 0$, but $V_0\ll\Omega$. The $\hat H_{eff}^{(\Omega)}$ results in $|\psi_R(t)\rangle=\left[\cos \frac{V_0t}{8}|gg\rangle+i\sin\frac{V_0 t}{8}|rr\rangle\right]\exp(-i3V_0t/8)$ and then projecting on to the rotated basis states given in Eqs. (\ref{rbs}), we get for $\omega=0$ and $V_0\ll\Omega$:
\begin{eqnarray}
P_{gg}(t)&=&\cos^4\frac{\Omega t}{2}\cos^2\frac{V_0t}{8}+\sin^4\frac{\Omega t}{2}\sin^2\frac{V_0t}{8} \\
P_{gr}(t)&=&P_{rg}(t)=\frac{1}{4}\sin^2 \Omega t \\
P_{rr}(t)&=&\cos^4\frac{\Omega t}{2}\sin^2\frac{V_0t}{8}+\sin^4\frac{\Omega t}{2}\cos^2\frac{V_0t}{8}.
\end{eqnarray}
These results are in excellent agreement with the numerical results obtained by solving the Schr\"odinger equation using the Hamiltonian in Eq. (\ref{ham3}) when $V_0\ll\Omega$ is satisfied. For both $\omega$ and $V_0$ are non zero but very small compared to $\Omega$, it is clear from the effective Hamiltonian in Eq. (\ref{heff3}) that there exists a competition between $V_0$ and $\omega$ in coupling the state $|gg\rangle$ to other states, as we have already discussed in Sec. \ref{dyn1}. With the initial $|gg\rangle$ state, the unitary evolution of $H_{eff}^{(\Omega)}$ results in,
\begin{widetext}
\begin{eqnarray}
\nonumber
    |\Psi_R(t)\rangle&=& e^{-iV_0 t/4}\times \left[ \frac{1}{2\eta}\left(\eta\left(\cos{\omega t/2}+ \cos{\eta t/4}\right)- iV_0\sin{\eta t/4}\right)|gg\rangle -\frac{i}{4\omega\eta}\left(\eta^2\sin{\eta t/4} +2\omega\eta \sin{\omega t/2}- V_0^2 \sin{\eta t/4}\right)|gr\rangle \right.\\ &&\left. +\frac{i}{4\omega\eta}\left(\eta^2\sin{\eta t/4}- 2\omega\eta\sin{\omega t/2}- V_0^2 \sin{\eta t/4}\right)|rg\rangle -\frac{1}{2\eta}\left(\eta\left(\cos{\eta t/4}- \cos{\omega t/2}\right)-iV_0\sin{\eta t/4}\right)|rr\rangle\right],
\end{eqnarray}
\end{widetext}
and then projecting to the rotated basis states given in Eqs. (\ref{rbs}) we get for $\{V_0, \omega\}\ll \Omega$
\begin{align}
    \begin{split}
        P_{gg}(t)=& \frac{1}{4}\left(\cos{\frac{(\omega+2\Omega)t}{2}}+ \cos{\frac{\eta t}{4}}\right)^2+ \frac{V_0^2}{4\eta^2}\sin^2{\frac{\eta t}{4}}\\
        P_{gr}(t)=& \left(\frac{1}{2}\sin{\frac{(\omega+2\Omega)t}{2}} + \frac{\omega}{\eta}\sin{\frac{\eta t}{4}}\right)^2\\
        P_{rg}(t)=& \left(\frac{1}{2}\sin{\frac{(\omega+2\Omega)t}{2}} - \frac{\omega}{\eta}\sin{\frac{\eta t}{4}}\right)^2\\
        P_{rr}(t)=& \frac{1}{4}\left(\cos{\frac{(\omega+2\Omega)t}{2}}- \cos{\frac{\eta t}{4}}\right)^2+ \frac{V_0^2}{4\eta^2}\sin^2{\frac{\eta t}{4}}
    \end{split}
\end{align}
where $\eta =\sqrt{V_0^2 +4\omega^2}$. The time average values become
\begin{equation}
\bar P_{gg}=\bar P_{rr}=1/4+V_0^2/(8\eta^2),
\label{prr}
\end{equation}
 and $\bar P_{gr}=\bar P_{rg}=1/8+\omega^2/(2\eta^2)$. When $\omega=0$, we retrieve the old results. For $\omega\neq 0$, $\bar P_{rr}$ is an increasing function of $V_0$ which explains why $\bar P_{rr}$ increases initially with $|V_0|$ at small values of $\omega$. The saturation point $V_0^P$ shown in Fig. \ref{fig:5}(a) for small $\omega$ can be obtained by equating the $\bar P_{rr}$ given in Eq. (\ref{prr}) to the Lorentzian profile,  $f(V_0/\Omega)= 3\Omega^2/8[\Omega^2 + (V_0/\nu_{rr})^2]$, obtained for $\bar P_{rr}$ with $\omega=0$, in which $\nu_{rr}$ is obtained by fitting to the exact numerical results. Doing so, we get
\begin{equation}
	V_0^p= \sqrt{\frac{2}{3}}\sqrt{\omega\left(-2\omega+ \sqrt{4\omega^2+ 3\nu_{rr}^2\Omega^2}\right)},
	\label{vop}
\end{equation}
which is shown as a dashed line in Fig. \ref{fig:5}(a). A comparison of time average populations from exact numerics, perturbation theory (Appendix \ref{a1}) and using the effective Hamiltonian is shown in Fig. \ref{fig:6} for small values of $\omega/\Omega$ and $V_0/\Omega$, and they are in good agreement with each other.

%%%%%
\section{Quantum Correlations}
\label{corr}
\begin{widetext}
\begin{figure*}
\centering
\includegraphics[width=1.75 \columnwidth]{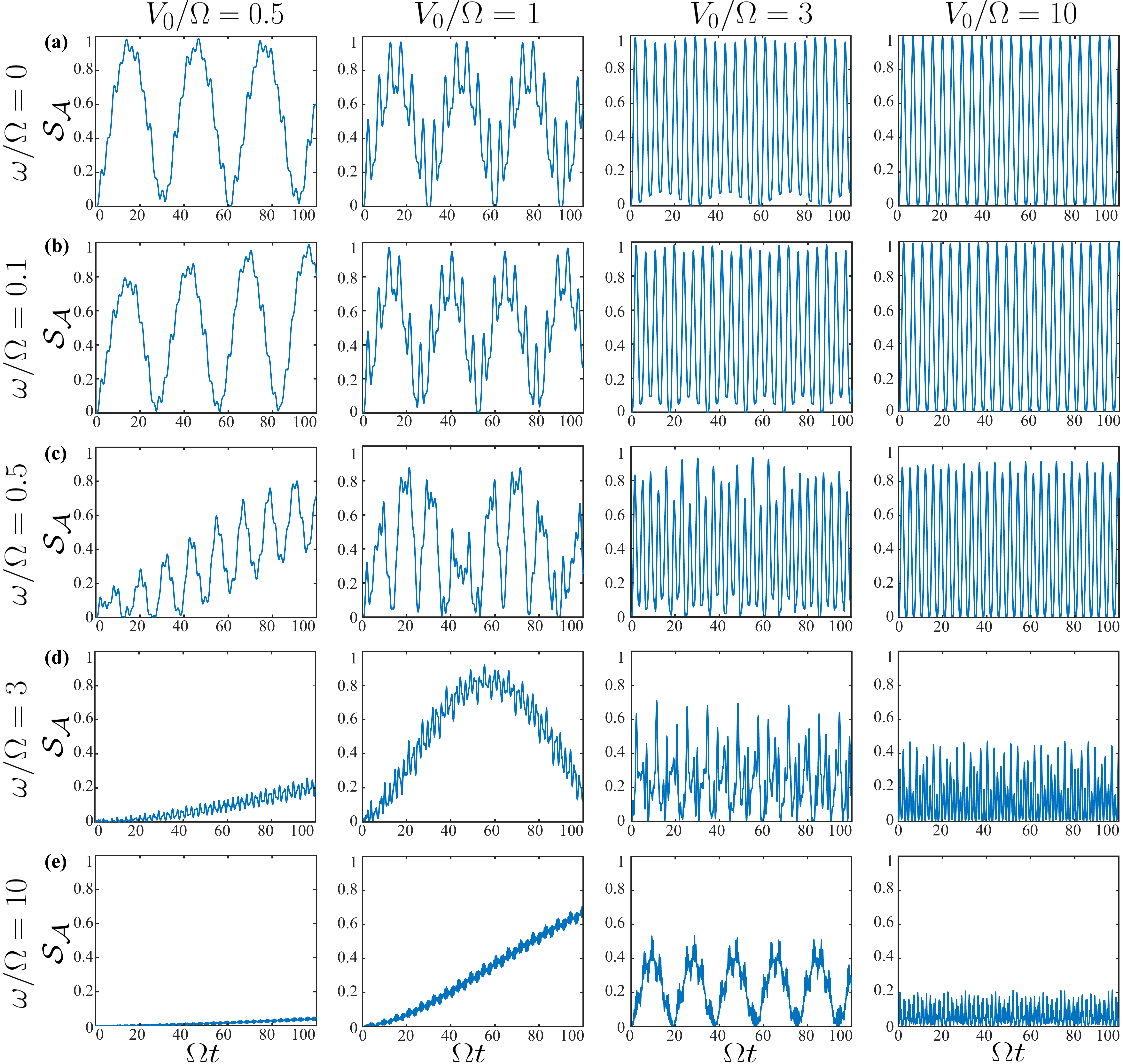}
\caption{(a) The time evolution of the entanglement entropy $\mathcal S_A(t)$ obtained from the reduced density matrix of the first atom for different $\omega/\Omega$ and $V_0/\Omega$ corresponding to the coherent dynamics discussed in Sec. \ref{dyn1}. In this case $\mathcal S_A$ is same as the discord $\mathcal D$. The values of $\omega/\Omega$ and $V_0/\Omega$ are indicated in the left and top sides respectively.}
\label{fig:7}
\end{figure*}
\end{widetext}
\begin{figure}
\centering
\includegraphics[width=1 \columnwidth]{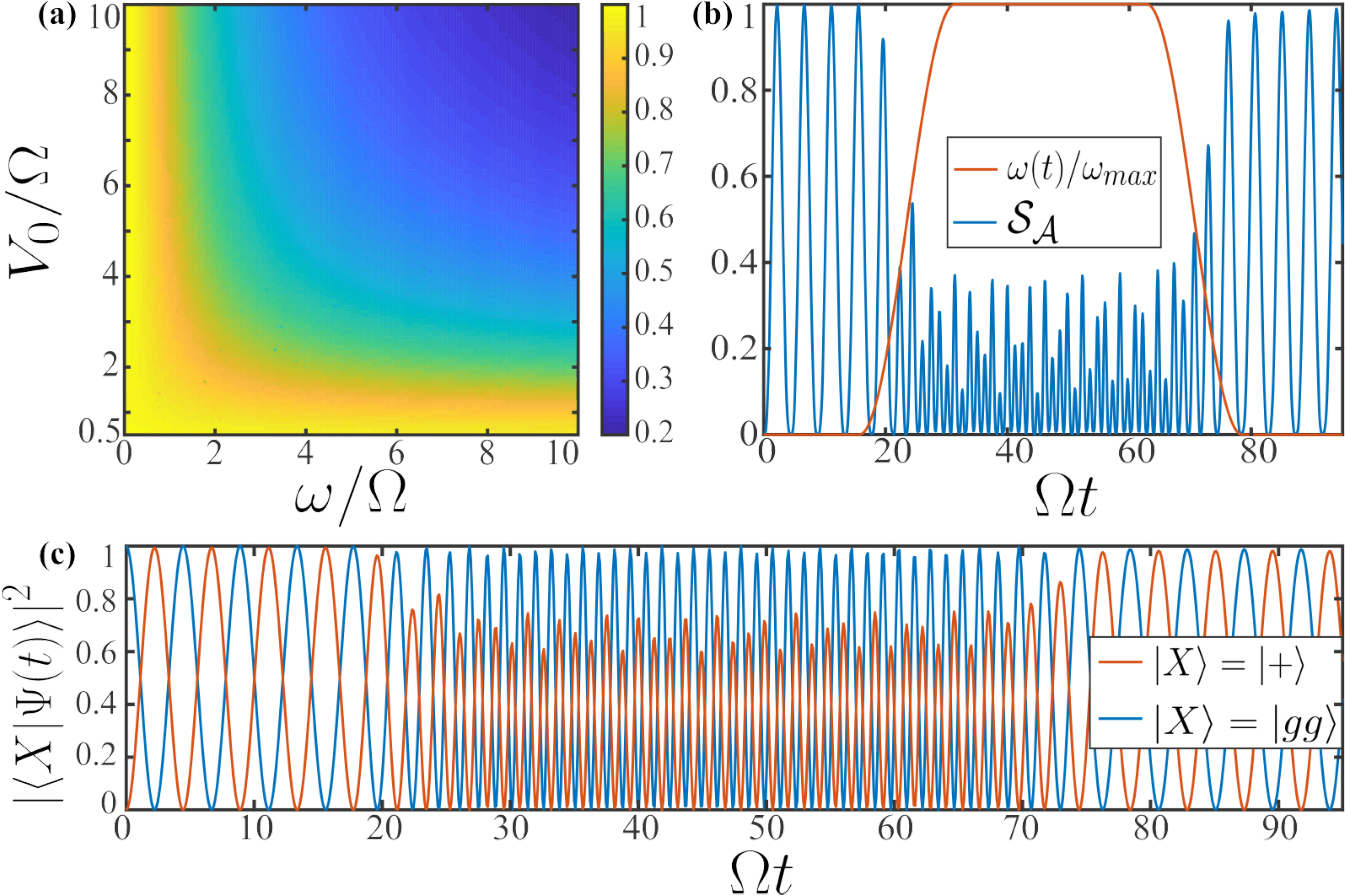}
\caption{(a) The maximum of $\mathcal S_A(t)$ as a function of $\omega/\Omega$ and $V_0/\Omega$ corresponding to the coherent dynamics discussed in Sec. \ref{dyn1}. The lower cutoff of $V_0/\Omega=0.5$ in the vertical axis is because for smaller values, the time taken to attain maximum $\mathcal S_A(t)$ becomes extremely large. (b) The dynamics of $\mathcal S_A(t)$ when the Rabi-offset $\omega$ is varied in time with $\alpha=0.1\Omega$ (see text), $V_0/\Omega=10$ and $\omega_{max}/\Omega= 4$. When $\omega$ reaches the maximum, $\mathcal S_A(t)$  is significantly suppressed, and is retrieved back to the initial dynamics once $\omega$ is brought back to zero. (c) shows the overlap of $|\psi(t)\rangle$ on the states $|gg\rangle$ and $|+\rangle$ for the dynamics shown in (b).}
\label{fig:8}
\end{figure}

In this section, we analyze the growth and subsequent evolution of the entanglement entropy $\mathcal S_A$ for the coherent dynamics discussed in Sec. \ref{dyn1}. At $t=0$, we have $\mathcal S_A(0)=0$ since the initial state $|gg\rangle$ is a separable state. Fig. \ref{fig:7} shows $\mathcal S_A(t)$ for different values of $V_0/\Omega$ and $\omega/\Omega$ and, is an oscillating function of time. Each row of four plots is for a fixed $\omega/\Omega$ and different $V_0/\Omega$, and vice versa for the columns.  For $V_0=0$, there is no correlation ($\mathcal S_A=0$) between the atoms. When $\omega=0$ and for any interaction strengths ($|V_0|\neq 0$), $\mathcal S_A(t)$ oscillates between 0 and its maximum possible value of $\log_22=1$. Larger the value of $V_0$, the maximum correlation is attained between shorter intervals of time. For sufficiently large $V_0$, $\mathcal S_A(t)$ exhibits clean periodic oscillation between 0 and 1, indicating a complete blockade in which the system exhibits coherent Rabi oscillations between the separable $|gg\rangle$ state and a maximally entangled $|+\rangle$ state. Making $\omega\neq 0$ significantly changes the growth and dynamics of the quantum correlations depending on the value of $V_0$. Not only $\omega$ slows down the correlation growth but also lower the maximum correlation that can be attained. When both $\omega$ and $V_0$ are very large compared to $\Omega$, the correlations are very well suppressed due to the Rydberg-biased freezing (see Fig. \ref{fig:7}). For sufficiently small  $V_0$, increasing $\omega$ slowed down the growth of $\mathcal S_A(t)$ but did not affect the maximum value of $\mathcal S_A$ attained over time (see along the first column in Fig. \ref{fig:7} and also Fig \ref{fig:8}(a) in which the maximum of $\mathcal S_A(t)$ is shown). It indicates that $\omega$ effectively reduces the effect of interaction strengths between the two atoms for small $V_0$ or in other words, there exists a competition between $V_0$ and $\omega$ as we pointed out earlier in Secs. \ref{dyn1} and \ref{heffs}. The maximum value of $\mathcal S_A(t)$ as a function of $V_0/\Omega$ and $\omega/\Omega$ is shown in Fig. \ref{fig:8}(a). We introduced a lower cutoff for $V_0$ in the vertical axis in Fig. \ref{fig:8}(a) since the time taken to attain the maximum correlation becomes extremely large for such small values of $V_0$ with large $\omega$. 
%%%%%%%%%

The above results opens up the possibility that the quantum correlations between the two atoms can be easily controlled by means of the Rabi-offset. To demonstrate that, we consider a time dependent $\omega$ [see Fig. \ref{fig:8}(b)] as follows: 
\begin{equation}
    \omega(t)/\omega_{max}= \begin{cases}
    0 , \ \ \ \   0\leq \alpha t\leq \pi/2 \\
\cos^2(\alpha t), \ \ \ \  \pi/2\leq \alpha t \leq \pi \\
1,  \ \ \ \ \pi \leq \alpha t \leq 2\pi  \\
\cos^2(\alpha t), \ \ \ \ 2\pi \leq \alpha t \leq 5\pi/2 \\
0, \ \ \ \  5\pi/2 \leq \alpha t \leq  3\pi\\
    \end{cases}
\end{equation}
where $\alpha$ determines the rate at which $\omega$ is varied. We take sufficiently large $V_0$ such that the two-atom
setup is in the fully blockade region thus, $\mathcal S_A(t)$  exhibits periodic oscillation between 0 and 1 in the absence of any Rabi-offset. Starting from the initial state $|gg\rangle$, first we slowly ramp $\omega(t)$ to $\omega_{max}$. The $\omega(t)$ suppresses the population in $|rg\rangle$ and consequently in $|+\rangle$ state [see Fig. \ref{fig:8}(c)]. The latter results in a significant loss of quantum correlations between the atoms. As we reduce $\omega(t)$ back to zero, the correlations are again build up in the system and completely retrieve its maximum value of 1 as $\omega$ vanishes. In the example shown in Figs. \ref{fig:8}(b) and (c), the value of $\alpha$ is taken such that not only the correlations are rebuild, but also the initial blockade dynamics is retrieved completely. The latter is verified by calculating the overlap functions $|\langle X|\psi(t)\rangle|^2$ where $|X\rangle\in \{|gg\rangle, |+\rangle\}$, see Fig. \ref{fig:8}(c). 

\vspace{1 cm}
\section{Dissipative dynamics}
\label{diss}
At this point, we discuss the effect of spontaneous emission rate from the Rydberg state $|r\rangle$ on the dynamics as well as the quantum correlations in the steady states of the master equation in Eq. (\ref{meq}). Since the dissipation drives the system into a mixed state, entanglement entropy is no longer a good measure for quantum correlations \cite{ben96, ved97, ved98}, and we restrict ourselves to quantum discords. As stated before, for a mixed state, $\mathcal D(A:B)$ may not be always equal to  $\mathcal D(B:A)$. In our two atoms setup, $\mathcal D(A:B)=\mathcal D(B:A)$ only when $\omega=0$, due to the exchange symmetry between the atoms.

\begin{figure}
\centering
\includegraphics[width=1 \columnwidth]{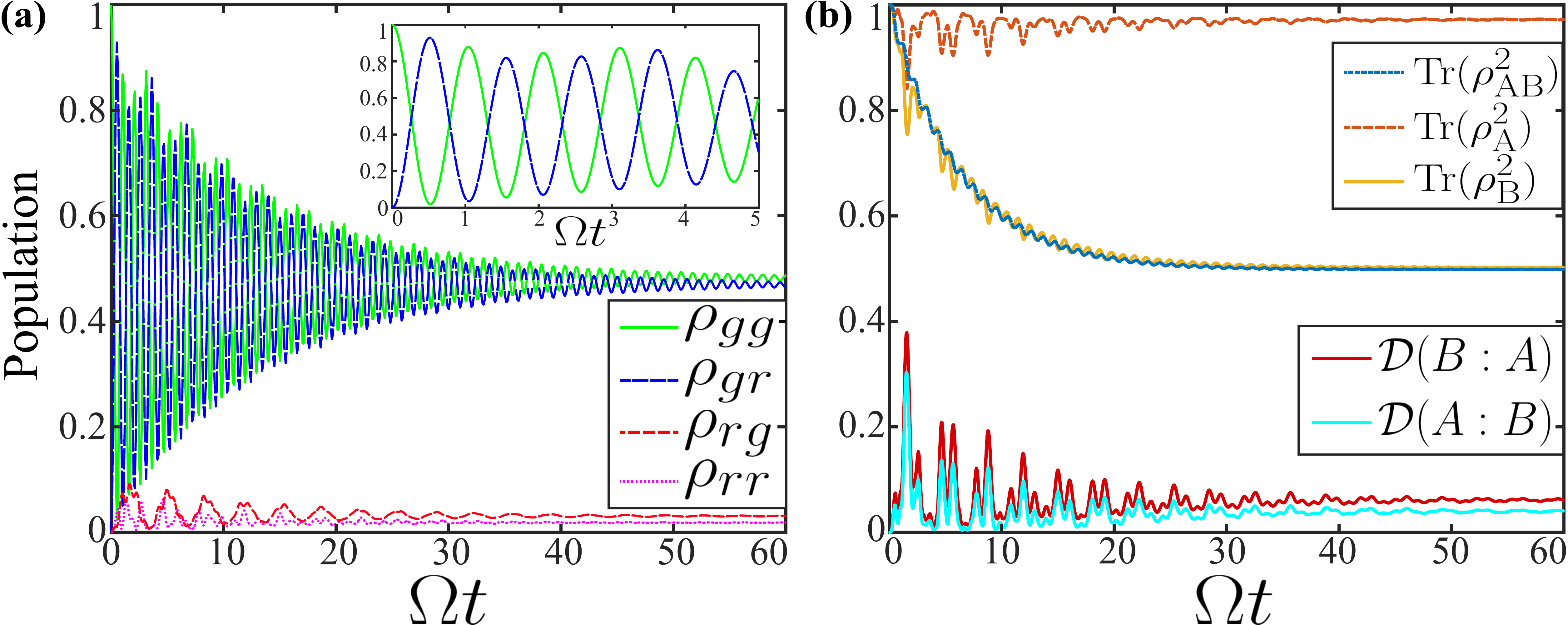}
\caption{(a) The populations vs time for $\omega/\Omega=V_0/\Omega=5$ and $\Gamma/\Omega=0.1$. The inset shows the same for the initial period of time. (b) shows the time evolution of both the quantum discords, and the purity of the total system and subsystems, for the dynamics shown in (a).}
\label{fig:9}
\end{figure}

\begin{figure}
\centering
\includegraphics[width=1 \columnwidth]{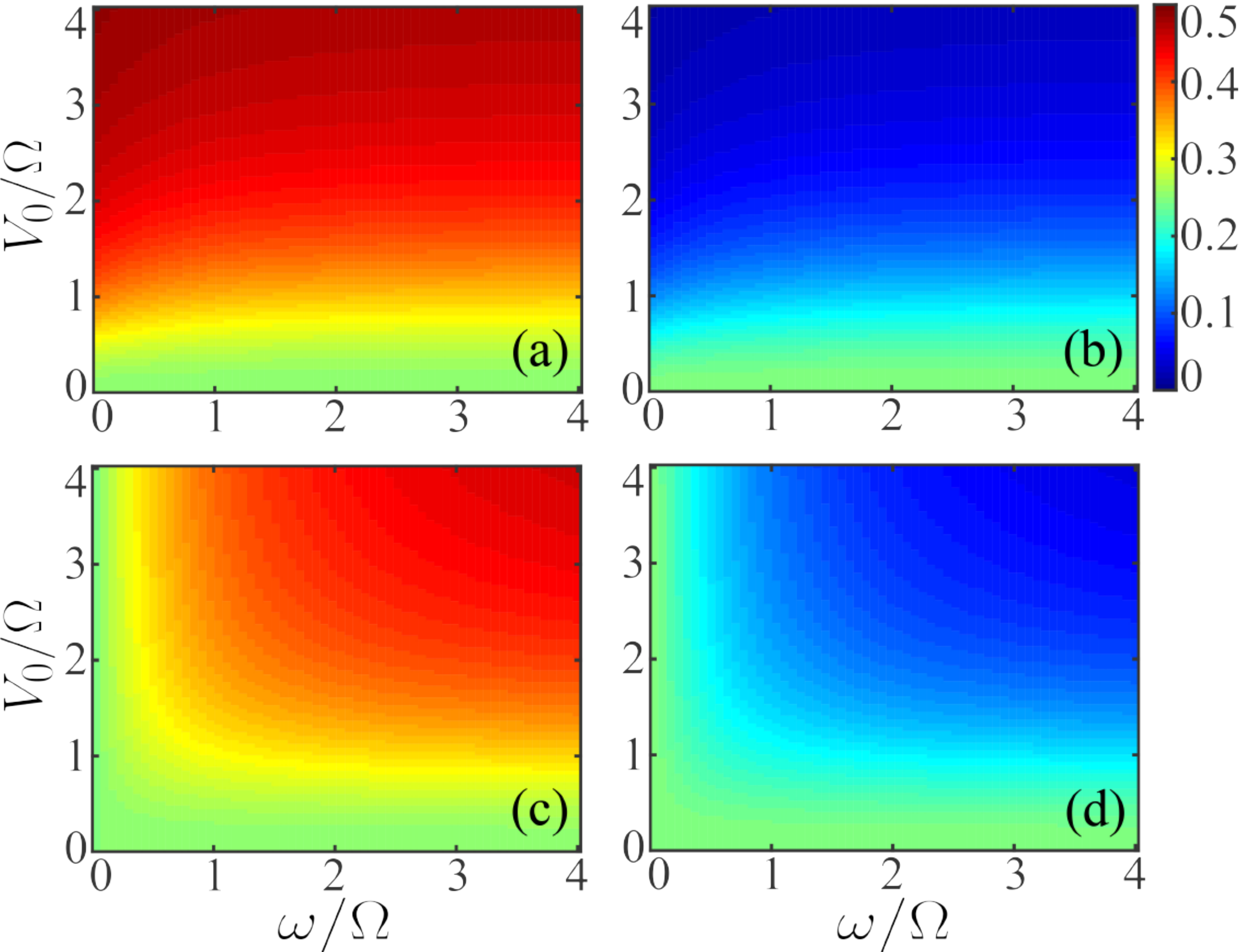}
\caption{The steady state populations: (a) $\rho_{gg}$, (b) $\rho_{rr}$, (c) $\rho_{gr}$ and (d) $\rho_{rg}$ as a function of $\omega/\Omega$ and $V_0/\Omega$ with $\Gamma/\Omega=0.1$. }
\label{fig:10}
\end{figure}

First, we look at the dynamics in the Rydberg-biased freezing regime by taking a large value for $\omega/\Omega$ and $V_0/\Omega$, see Fig. \ref{fig:9}. In Fig \ref{fig:9}(a) we show the time evolution of the populations [$\rho_{\alpha\beta}(t)$ with $\alpha, \beta\in \{g, r\}$] for $\omega/\Omega=V_0/\Omega=5$ and $\Gamma/\Omega=0.1$. At shorter times we see the damped Rabi oscillations between the states $|gg\rangle$ and $|gr\rangle$ [inset of Fig \ref{fig:9}(a)] whereas the states $|rg\rangle$ and $|rr\rangle$ are almost suppressed. Eventually, the system reaches the steady state with almost equal populations between $|gg\rangle$ and $|gr\rangle$. Fig \ref{fig:9}(b) shows both the correlations and the trace of the square of the density matrices of both the total system and the subsystems as a function of time, for the dynamics shown in Fig \ref{fig:9}(a). The quantity $\rm{Tr}\left(\hat\rho^2_{AB}\right)$ measures the purity of the total system, and  $\rm{Tr}\left(\hat\rho^2_{A}\right)$ $\left[\rm{Tr}\left(\hat\rho^2_{B}\right)\right]$ measures that of the subsystem $A$ [$B$]. Note that, $\rm{Tr}\left(\hat\rho^2_{A}\right)$ remains close to unity during the dissipative evolution with small fluctuations initially, and becomes steady at unity as the system converges to the steady state. This makes sense, because the first atom remains frozen in the ground state due to the Rydberg-biased freezing and thus, in a pure state. Whereas, for the second atom (subsystem $B$), $\rm{Tr}\left(\hat\rho^2_{B}\right)$ decreases and eventually converges to 1/2 indicating that it is in a completely mixed state i.e., a mixture of $|g\rangle$ and $|e\rangle$ with equal populations.  As a consequence, the density matrix of the whole system is not completely mixed, and  $\rm{Tr}\left(\hat\rho^2_{AB}\right)$ converges to 1/2. The evolution of the corresponding quantum discords $\mathcal D(A:B)$ and $\mathcal D(B:A)$ are shown in Fig \ref{fig:9}(b). Since the initial state is $\rho_{gg}=1$, a pure product state, we have vanishing discords at $t=0$. In the initial period of time, they exhibit non-periodic oscillations and converge to a small value due to the Rydberg-biased freezing, as the system approaches steady state. We have $\mathcal D(B:A)>\mathcal D(A:B)$ since the purity of $\hat \rho_A$ remains larger than $\hat\rho_B$ at any instant.

\begin{figure}
\centering
\includegraphics[width=1 \columnwidth]{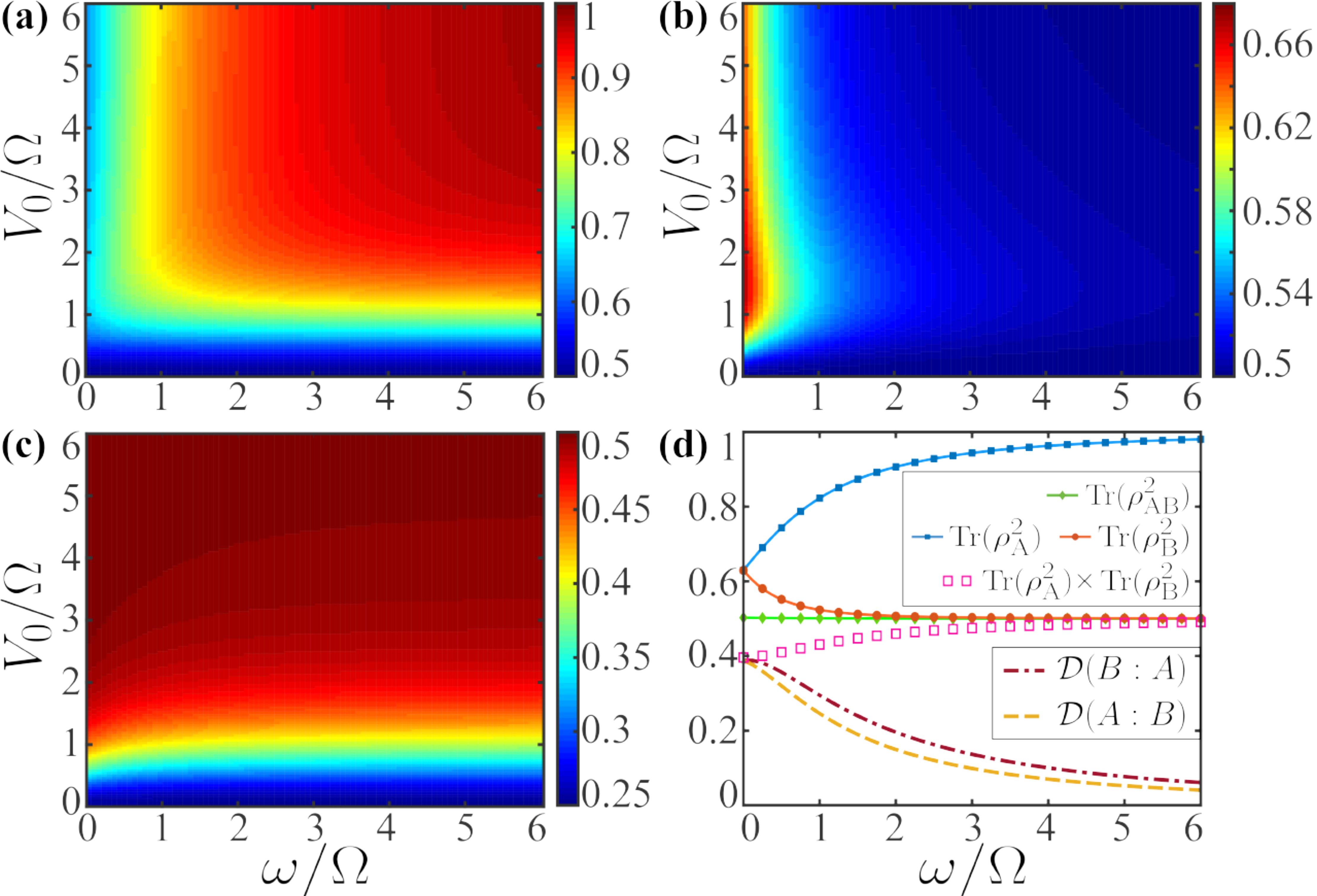}
\caption{The  steady state purity of the subsystems (a)  $A$ and  (b) $B$, and (c) the full system ($AB$) as a function of $\omega/\Omega$ and $V_0/\Omega$.  (d) shows the purity and the quantum discords as a function of $\omega/\Omega$ for $V_0/\Omega=10$ (blockade region).  $\Gamma/\Omega=0.1$ for all figures. In (d) the solid lines are the analytical results given in Appendix \ref{sss} and points are from the full numerical calculations for the steady state purity. Dashed lines show the quantum discords. The open squares show $\rm{Tr}\left(\hat\rho^2_{A}\right)\times\rm{Tr}\left(\hat\rho^2_{B}\right)$ which matches to $\rm{Tr}\left(\hat\rho^2_{AB}\right)$ at large $\omega$.}
\label{Fig:11}
\end{figure}

\begin{figure}
\centering
\includegraphics[width=1 \columnwidth]{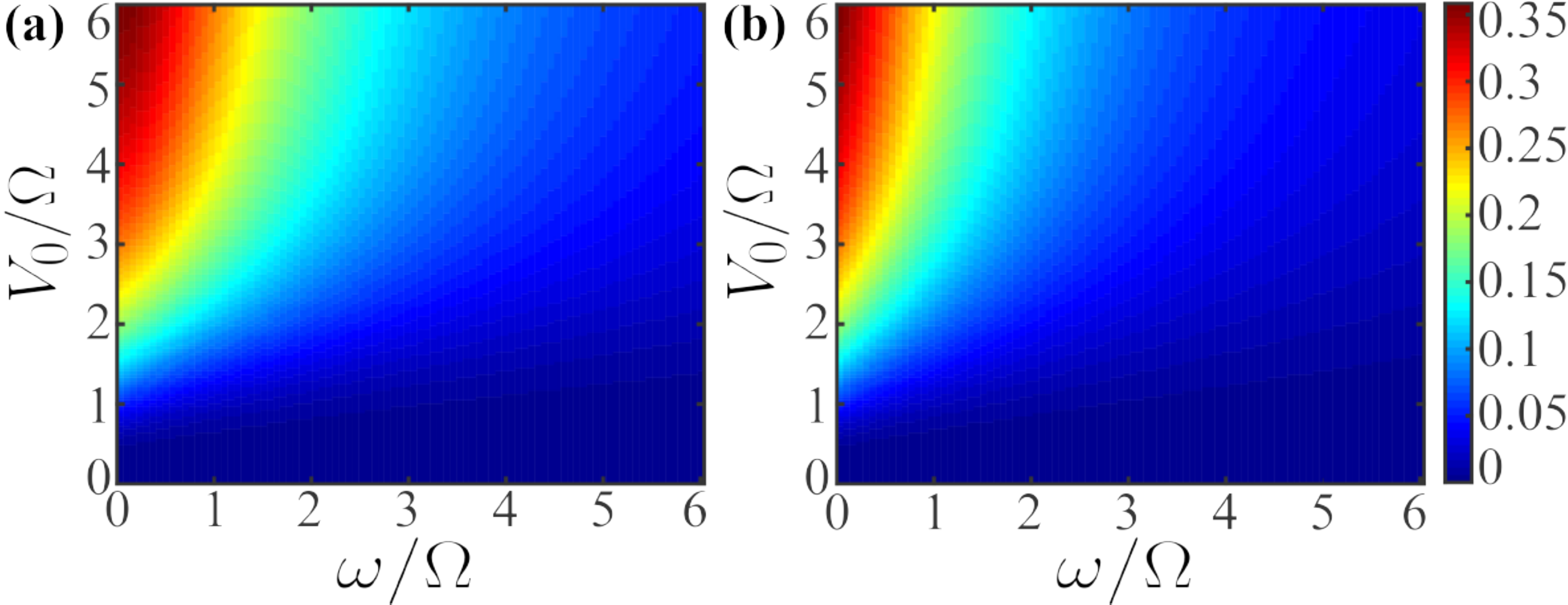}
\caption{The steady state quantum discords (a) $\mathcal D(B:A)$  and (b) $\mathcal D(A:B)$  as a function of $\omega/\Omega$ and $V_0/\Omega$ for $\Gamma/\Omega=0.1$. For $\omega\neq 0$ we have  $\mathcal D(B:A)\neq\mathcal D(A:B)$. For $\omega=0$, the correlations maximally saturate at large $V_0$ due to Blockade, and they start to diminish as $\omega$ increases. Discords vanish in the Rydberg-biased freezing regime where the system is described by a product state. }
\label{Fig:12}
\end{figure}

Further, we extend our calculations to wider range of $V_0/\Omega$ and $\omega/\Omega$. In Fig. \ref{fig:10} we show the steady state populations: $\rho_{gg}$, $\rho_{gr}$, $\rho_{rg}$, and $\rho_{rr}$ as a function of $V_0/\Omega$ and $\omega/\Omega$ with $\Gamma/\Omega=0.1$. Two features are evident from the Fig. \ref{fig:10}(b): (i) the doubly excited state ($\rho_{rr}$) is completely suppressed at large $V_0$ due to the blockade, and (ii) for sufficiently large $\omega$ the blockade criteria is independent of $\omega$ similar to that in the case of coherent dynamics. At sufficiently large values of $V_0$ and $\omega$, both $\rho_{rr}$ [Fig. \ref{fig:10}(b)] and $\rho_{rg}$ [Fig. \ref{fig:10}(d)] approaches zero and the populations are shared among $\rho_{gg}$ [Fig. \ref{fig:10}(a)] and $\rho_{gr}$ [Fig. \ref{fig:10}(c)], as discussed above. More insights into the steady states are attained from the purity of the system and the subsystems [Fig. \ref{Fig:11}]. For small values of $V_0 (\ll \Omega)$, independently the value of $\omega$, at the steady state, the total system [$\rm{Tr}\left(\hat\rho^2_{AB}\right)\sim 0.25$] as well as the subsystems [$\rm{Tr}\left(\hat\rho^2_{A}\right)\sim 0.5$  and $\rm{Tr}\left(\hat\rho^2_{B}\right)\sim 0.5$] are completely mixed with no quantum correlations between the subsystems, see Fig. \ref{Fig:12} for the corresponding discords. Thus, the steady state of the system is a product state, $\rho_{AB}=\rho_A\otimes\rho_B$ and consequently $\rm{Tr}\left(\hat\rho^2_{AB}\right)=\rm{Tr}\left(\hat\rho^2_{A}\right)\times\rm{Tr}\left(\hat\rho^2_{B}\right)$. This also indicates that for sufficiently small $V_0$, the correlations which initially build up in the system from the interactions have been washed out eventually by the dissipation.  As $V_0$ increases, for sufficiently small $\omega$, the correlations survive in the steady state as shown in Fig. \ref{Fig:12}, and their magnitude increases with increase in $V_0$, and eventually saturates to a constant value ($\Gamma$-dependent) at large values of $V_0(\sim 10 \Omega)$. The strong correlations at large $V_0$ for small $\omega$ is attributed to the Rydberg blockade \cite{fan17}. The maximally saturated correlation attained at large $V_0$ for a given Rabi-offset decreases with increase in $\omega$ as shown in Fig. \ref{Fig:11}(d). 

 Concerning the purity, as $V_0$ increases $\rm{Tr}\left(\hat\rho^2_{AB}\right)$ increases and saturates to a value of 0.5. Interestingly, the latter happens independent of the value of $\omega$ [Fig. \ref{Fig:11}(c)], but a complete picture is accessible only through $\rm{Tr}\left(\rho^2_{A}\right)$ and $\rm{Tr}\left(\rho^2_{B}\right)$. For instance, the purity as a function of $\omega/\Omega$ in the blockade region ($V_0/\Omega=10$) is shown in Fig. \ref{Fig:11}(d). Though, $\rm{Tr}\left(\rho^2_{AB}\right)\sim 0.5$ is independent of $\omega$ at large $V_0$, the purity of the subsystems depends strongly on $\omega/\Omega$. The purity in the strongly driven atom (subsystem $B$) decreases as a function of $\omega/\Omega$ and becomes maximally mixed at sufficiently large values, whereas that of first atom (subsystem $A$) increases with $\omega$ and eventually becomes a pure state at very large values of $\omega/\Omega$. Thus, the total system is in a product state for large $V_0$ and $\omega$, and the purity of the  becomes $\rm{Tr}\left(\hat\rho^2_{AB}\right)=\rm{Tr}\left(\hat\rho^2_{A}\right)\times\rm{Tr}\left(\hat\rho^2_{B}\right)$ as shown in Fig. \ref{Fig:11}(d). That means, the quantum correlations decrease with increase in $\omega$ for large $V_0$, see Fig. \ref{Fig:11}(d) for the corresponding quantum discords. Note that, we also obtained the analytical results for steady state density matrices, the purity of the system and subsystems, see Appendix \ref{sss}, and are in excellent agreement with the exact numerical calculations shown in Fig. \ref{Fig:11}.

%%%%%%%%%%%%%%
\section{conclusions}
In conclusion, we studied  the dynamics and the quantum correlations in a minimal setup of two two-level Rydberg atoms driven continuously and independently by two distinct laser fields. In particular, we analyzed the effect of an offset in Rabi frequencies between the fields on the Rydberg excitation dynamics, in the presence of Rydberg-Rydberg interactions. Interestingly, we identified novel features in the system, amplifying the driving in one atoms freezes the dynamics in the second atom, in the blockade regime. Non-monotonous behaviour of doubly excited state population as a function interaction strength for small Rabi-offsets is observed. The effective Hamiltonians obtained via unitary transformations at various limits of system parameters provide us analytical solutions for the dynamics. They are found to be in excellent agreement with the complete numerical calculations. The quantum correlations for both the coherent and dissipative dynamics are studied. We also demonstrate that the quantum correlations can be controlled using time dependent Rabi-offset. Our studies open up a new possibility to engineer quantum states in Rydberg atom setup, and the immediate question would be to extend the studies to more than two atoms, and also at different geometries. 

\section{Acknowledgments}
We acknowledge the funding from the Indo-French Centre for the Promotion of Advanced Research and UKIERI-UGC Thematic Partnership No. IND/CONT/G/16-17/ 73 UKIERI-UGC project.
%%%%%%%%%%%%%%%%%%%
\appendix
\section{Perturbation theory in the weak interaction limit}	
\label{a1}
We write the Hamiltonian in Eq. (\ref{ham3})  as $\hat{H} = \hat{H}_{C} + \hat{H}_{V}$ where $\hat{H}_{C}=(\Omega/2)\sum_{i=1}^2\hat\sigma_x^{i}+(\omega/2)\hat\sigma_x^2$ and $\hat{H}_{V}=V_0\hat\sigma_{rr}^{1}\hat\sigma_{rr}^{2}$. Treating $\hat H_V$ as a perturbation, we obtain the perturbative corrections to the eigenvectors  and eigenvalues of $\hat{H}_{C}$ using non degenerate perturbation theory which demands $\omega\neq 0$. The eigenvalues of the unperturbed Hamiltonian $\hat H_{C}$ and the eigenvectors in the basis $\{|gg\rangle, |gr\rangle, |rg\rangle, |rr\rangle\}$: 

	\begin{equation}
	E_1^0= \frac{-(2\Omega + \omega)}{2}, \hspace*{1cm} |\Psi_1^0\rangle= \frac{1}{2}
	\left(\begin{matrix}
	1\\
	-1\\
	-1\\
	1
	\end{matrix}\right)
	\end{equation}
	\begin{equation}
	E_2^0= \frac{\omega}{2}, \hspace*{1cm} |\Psi_2^0\rangle= \frac{1}{2}
	\left(\begin{matrix}
	-1\\
	-1\\
	1\\
	1
	\end{matrix}\right)
	\end{equation}
	\begin{equation}
	E_3^0= \frac{-\omega}{2}, \hspace*{1cm} |\Psi_3^0\rangle= \frac{1}{2}
	\left(\begin{matrix}
	-1\\
	1\\
	-1\\
	1
	\end{matrix}\right)
	\end{equation}
	\begin{equation}
	E_4^0= \frac{2\Omega + \omega}{2}, \hspace*{1cm} |\Psi_4^0\rangle=  \frac{1}{2}
	\left(\begin{matrix}
	1\\
	1\\
	1\\
	1
	\end{matrix}\right)
	\end{equation}
	%%%%%%
			\begin{figure}
\centering
\includegraphics[width= .75\columnwidth]{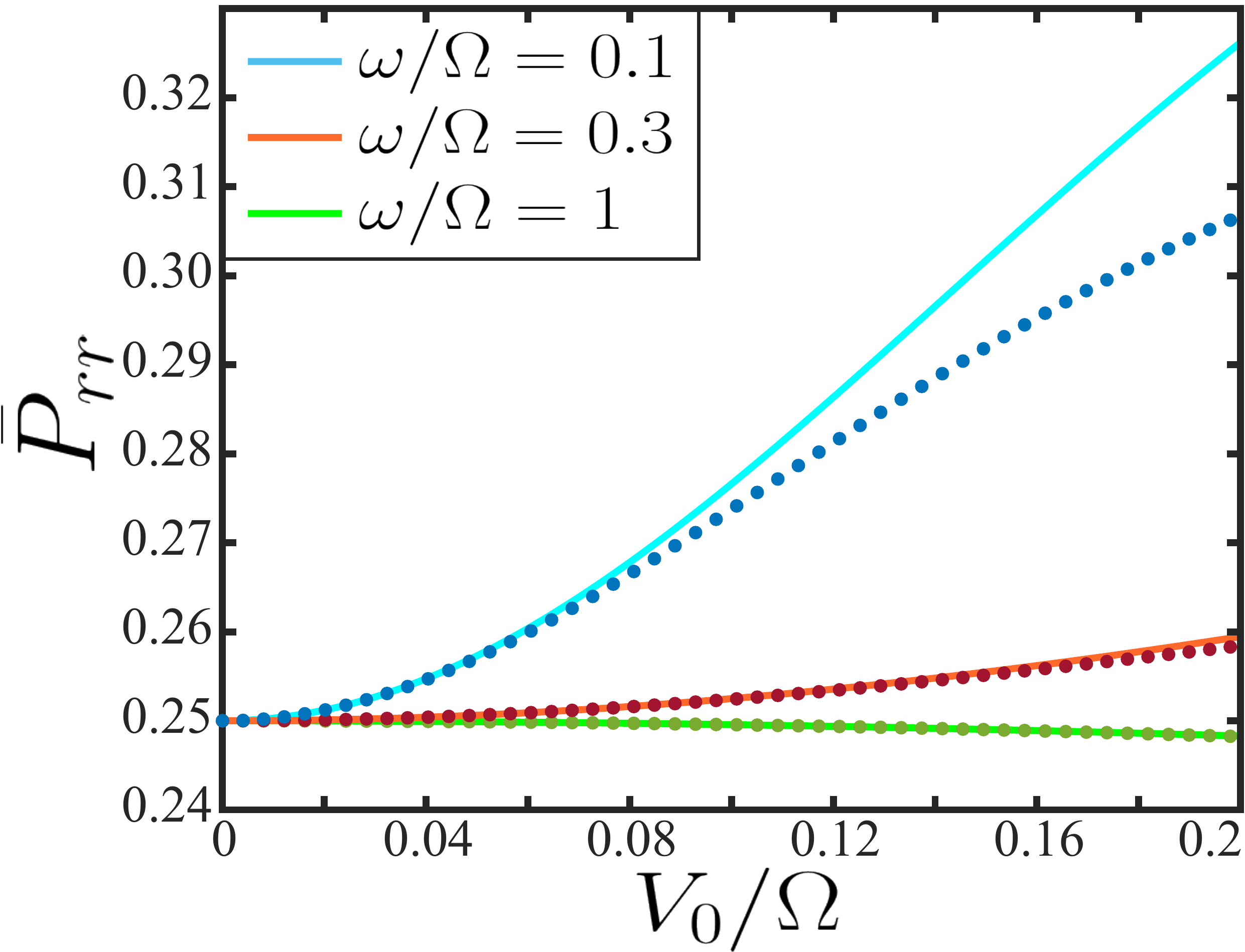}
\caption{The time average population in $|rr\rangle$ state obtained using the perturbation theory (solid lines) in the weak interaction limit and the exact numerical solution (dotted lines) as a function of $V_0$ for different $\omega$. The initial increment of $\bar P_{rr}$ in $V_0/\Omega$ for small $\omega/\Omega$, discussed in Sec. \ref{dyn1} is well captured by the second order perturbation theory.}
\label{figa1}
\end{figure}
	
	The first order correction to all the eigenvalues is simply $E_i^{1}=\langle\Psi_i^0|\hat H_V|\Psi_i^0\rangle=V_0/4$, and we get the second-order corrections as:
	\begin{align}
	    E_1^2=& -V_0^2 \left(\frac{(2\Omega+\omega)^2 + \Omega(\Omega+\omega)}{16\Omega(\Omega+\omega)(2\Omega+\omega)}\right)=-E_4^2\\
	    E_2^2=& -V_0^2 \left(\frac{\omega^2-\Omega(\Omega+\omega)}{16\Omega\omega(\Omega+\omega)}\right)=-E_3^2\\   \nonumber	\end{align}
The first order correction to the eigenstates is obtained by 
	\begin{equation}
	|\Psi_i^1\rangle = \sum_{j \neq i} \frac{\bra{\Psi_j^0}H_V\ket{\Psi_i^0}}{E_i^0 - E_j^0} |\Psi_j^0\rangle,
	\end{equation}
	which then gives us,
	\begin{equation}
	|\Psi_1^1\rangle = \frac{-V_0/8}{(\Omega+\omega)} 
	\left(\begin{matrix}
	-1\\
	-1\\
	1\\
	1
	\end{matrix}\right) 
	- \frac{V_0/8}{\Omega} 
	\left(\begin{matrix}
	-1\\
	1\\
	-1\\
	1
	\end{matrix}\right) 
	- \frac{V_0/8}{(2\Omega+\omega)} 
	\left(\begin{matrix}
	1\\
	1\\
	1\\
	1
	\end{matrix}\right).
	\end{equation}
	Thus, we have up to first order correction in $V_0$, the first eigenvector:
	\begin{equation}
	|\Psi_1\rangle = |\Psi_1^0\rangle + |\Psi_1^1\rangle = \frac{1}{2}
	\left(\begin{matrix}
	\frac{1}{4}\left(\frac{V_0}{\Omega+\omega}+ \frac{V_0}{\Omega} - \frac{V_0}{2\Omega+\omega}\right) + 1\\
	\frac{1}{4}\left(\frac{V_0}{\Omega+\omega}- \frac{V_0}{\Omega} - \frac{V_0}{2\Omega+\omega}\right)- 1\\
	\frac{1}{4}\left(-\frac{V_0}{\Omega+\omega}+ \frac{V_0}{\Omega}- \frac{V_0}{2\Omega+\omega}\right)- 1\\
	-\frac{1}{4}\left(\frac{V_0}{\Omega+\omega}+\frac{V_0}{\Omega}+\frac{V_0}{2\Omega+\omega}\right)+ 1
	\end{matrix}\right)
	\end{equation} 
	
	%%%%
	
	%%%
	Similarly, the remaining eigenvectors are obtained as  
	\begin{equation}
	|\Psi_2 \rangle= |\Psi_2^0\rangle + |\Psi_2^1\rangle = \frac{1}{2}
	\left(\begin{matrix}
	\frac{1}{4}\left(\frac{V_0}{\Omega+\omega}- \frac{V_0}{\omega} - \frac{V_0}{\Omega}\right) - 1\\
	\frac{1}{4}\left(-\frac{V_0}{\Omega+\omega}+ \frac{V_0}{\omega} - \frac{V_0}{\Omega}\right)- 1\\
	-\frac{1}{4}\left(\frac{V_0}{\Omega+\omega}+\frac{V_0}{\omega}+\frac{V_0}{\Omega}\right)+ 1\\
	\frac{1}{4}\left(\frac{V_0}{\Omega+\omega}+ \frac{V_0}{\omega} - \frac{V_0}{\Omega}\right)+ 1
	\end{matrix}\right)
	\end{equation} 
	\begin{equation}
	|\Psi_3\rangle = |\Psi_3^0\rangle + |\Psi_3^1\rangle = \frac{1}{2}
	\left(\begin{matrix}
	\frac{1}{4}\left(\frac{V_0}{\Omega}+\frac{V_0}{\omega} - \frac{V_0}{\Omega+\omega}\right) - 1\\
	-\frac{1}{4}\left(\frac{V_0}{\Omega}-\frac{V_0}{\omega} + \frac{V_0}{\Omega+\omega}\right)+ 1\\
	-\frac{1}{4}\left(\frac{V_0}{\Omega}+\frac{V_0}{\omega}+\frac{V_0}{\Omega+\omega}\right)- 1\\
	\frac{1}{4}\left(\frac{V_0}{\Omega}-\frac{V_0}{\omega}-\frac{V_0}{\Omega+\omega}\right)+ 1
	\end{matrix}\right)
	\end{equation}
	\begin{equation}
	|\Psi_4\rangle = |\Psi_4^0\rangle + |\Psi_4^1\rangle = \frac{1}{2}
	\left(\begin{matrix}
	\frac{1}{4}\left(\frac{V_0}{2\Omega+\omega}- \frac{V_0}{\Omega} - \frac{V_0}{\Omega+\omega}\right) + 1\\
	\frac{1}{4}\left(-\frac{V_0}{2\Omega+\omega}- \frac{V_0}{\Omega} + \frac{V_0}{\Omega+\omega}\right)+ 1\\	
	\frac{1}{4}\left(-\frac{V_0}{2\Omega+\omega}+ \frac{V_0}{\Omega} - \frac{V_0}{\Omega+\omega}\right)+ 1\\	
	\frac{1}{4}\left(\frac{V_0}{2\Omega+\omega}+ \frac{V_0}{\Omega} + \frac{V_0}{\Omega+\omega}\right)+ 1
	\end{matrix}\right)
	\end{equation} 	
	Then, writing the general time dependent solution as: 
	\begin{equation}
	|\Psi(t)\rangle= Ae^{-iE_1t} \ket{\Psi_1}+ Be^{-iE_2t}\ket{\Psi_2} + Ce^{-iE_3t}\ket{\Psi_3} + De^{-iE_4t}\ket{\Psi_4},
	\end{equation}
	with $E_i=E_i^0+E_i^1+E_i^2$ and $|\Psi(t=0)\rangle=|gg\rangle$. Using the initial condition and solving a set of coupled linear equations we get the expressions for the co-efficients:
	
	\begin{widetext}
	 \begin{eqnarray*}
	     A&=& \frac{2\Omega (\Omega+\omega)(2\Omega+\omega)}{\Lambda}\left[64\omega^2\Omega^3\left(\Omega+\omega\right)^3(2\Omega+\omega) + 16V_0 \omega^2\Omega^2(\Omega+\omega)^2\left(\omega^2+ 3\omega\Omega+ 3\Omega^2\right)+ 4V_0^2 \Omega(\Omega+\omega)(2\Omega+\omega)\times \right.\\ && \left. \left(\omega^4 + 4\omega^3\Omega+5\omega^2\Omega^2+2\omega\Omega^3+\Omega^4\right)+V_0^3(\Omega^6 - 3\Omega^5 (\Omega+\omega) + 3\Omega^3 (\Omega+\omega)^3 - 3\Omega (\Omega+\omega)^5 + (\Omega+\omega)^6)\right]\\
	      B&=& \frac{-2\Omega\omega (\Omega+\omega)}{\Lambda}\left[64\omega\Omega^3(\Omega+\omega)^3(2\Omega+\omega)^2 + 16V_0\Omega^2(\Omega+\omega)^2 (2\Omega+\omega)^2\left(\omega^2 +\omega\Omega+\Omega^2\right)+4V_0^2 \omega\Omega(\Omega+\omega)\times \right. \\ && \left. \left(\omega^4 + 4\omega^3\Omega + 5\omega^2\Omega^2 +2\omega\Omega^3 +\Omega^4\right)+ V_0^3\left(\Omega^6 + 3\Omega^5 (\Omega+\omega) - 3\Omega^3 (\Omega+\omega)^3 + 3\Omega (\Omega+\omega)^5 + (\Omega+\omega)^6\right) \right] \\
	       C&=& \frac{2\Omega\omega (\Omega+\omega)}{\Lambda}\left[-64\omega\Omega^3 (\Omega+\omega)^3 (2\Omega+\omega)^2+ 16V_0 \Omega^2(\Omega+\omega)^2 (2\Omega+\omega)^2\left(\omega^2+\omega\Omega+\Omega^2\right)-4V_0^2 \omega\Omega(\Omega+\omega)\times \right. \\ && \left.\left(\omega^4 +4\omega^3\Omega + 5\omega^2\Omega^2 +2\omega\Omega^3 +\Omega^4\right)+V_0^3\left(\Omega^6 + 3\Omega^5 (\Omega+\omega) - 3\Omega^3 (\Omega+\omega)^3 + 3\Omega (\Omega+\omega)^5 + (\Omega+\omega)^6\right)\right]\\
	        D&=& \frac{2\Omega (\Omega+\omega)(2\Omega+\omega)}{\Lambda}\left[64\omega^2\Omega^3 (\Omega+\omega)^3(2\Omega+\omega)-16V_0\omega^2\Omega^2(\Omega+\omega)^2\left(\omega^2+3\omega\Omega+ 3\Omega^2\right)+4V_0^2\Omega(\Omega+\omega)(2\Omega+\omega)\times \right. \\ &&\left.\left(\omega^4 + 4\omega^3\Omega+ 5\omega^2\Omega^2+2\omega\Omega^3 +\Omega^4\right)+V_0^3\left(-\Omega^6 + 3\Omega^5 (\Omega+\omega) - 3\Omega^3 (\Omega+\omega)^3 + 3\Omega (\Omega+\omega)^5 - (\Omega+\omega)^6\right)\right],
	 \end{eqnarray*}
	 \end{widetext}
	% \vspace {0.5 cm}
	where $\Lambda = 256\omega^2\Omega^4(\Omega+\omega)^4(2\Omega+\omega)^2+ 32V_0^2\Omega^2(\Omega+\omega)^2(\Omega^4+(\Omega+\omega)^4)+ \frac{V_0^4}{2}\left(\Omega^8+(\Omega+\omega)^8+\omega^4(2\Omega+\omega)^4\right)$. From above expressions it is straight forward to obtain $P_{rr}(t)$, and finally we have the time average population in the $|rr\rangle$ state as, 
\begin{equation}
\bar P_{rr}=A^2f_-^2(\omega, \Omega)+B^2g_+^2(\omega, \Omega)+C^2g_-^2(\omega, \Omega)+D^2f_+^2(\omega, \Omega)
\end{equation}
where the functions
\begin{align}f_\pm(\omega, \Omega)=& \frac{1}{2}\pm \frac{V_0}{8}\left[\frac{(2\Omega+\omega)^2 + \Omega(\Omega+\omega)}{\Omega(\Omega+\omega)(2\Omega+ \omega)}\right]\\
	g_\pm(\omega, \Omega)=& \frac{1}{2}\pm\frac{V_0}{8}\left[\frac{\Omega( \Omega+\omega)-\omega^2}{\Omega\omega(\Omega+\omega)}\right].
	\end{align}
	The perturbation theory results comparing that of exact numerical solutions are shown in Fig. \ref{figa1}.
		
%%%%%%%	

	\section{Analytical results for the steady state density matrices and purity of the system and subsystems}
	\label{sss}
	On solving $\dot \rho(t) = 0$, the steady state density matrix of the system is obtained as:
	\begin{widetext}
\begin{eqnarray*}
    \rho_{AB} &=& 
    \frac{1}{\kappa}\left[\left(V_0^2 \left(\Gamma^2 + \Omega^2 + (\Omega+ \omega)^2\right)+ (\Gamma^2 + \Omega^2)\left(\Gamma^2 + (\Omega+\omega)^2\right)\right)|gg\rangle\langle gg|+ \left(V_0^2+ \Gamma^2 + \Omega^2\right)\left(\Omega+\omega\right)^2|gr\rangle\langle gr|\right.\\&& \left. +\Omega^2\left(V_0^2 + \Gamma^2 + (\Omega+\omega)^2\right)|rg\rangle \langle rg| + \Omega^2 (\Omega+\omega)^2|rr\rangle\langle rr| \right.\\&& \left. +\Omega\left(V_0^2 + \Gamma^2\right)(\Omega+\omega)\left(|gr\rangle \langle rg| + \rm{H.c.}\right) + \left((\Omega+\omega)(i\Gamma- V_0)\left(\Gamma^2 + \Omega^2 - iV_0\Gamma\right)|gg\rangle\langle gr| +\rm{H.c.}\right)\right.\\&& \left. + \left(\Omega(-V_0 + i\Gamma)\left(\Gamma^2 + (\Omega+\omega)^2 -iV_0\Gamma\right)|gg\rangle\langle rg| + \rm{H.c.}\right)+ \left(-\Gamma(iV_0+\Gamma)\Omega(\Omega+\omega)|gg\rangle\langle rr|+\rm{H.c.}\right)\right.\\&& \left. + \left((-V_0+ i\Gamma)\Omega(\Omega+\omega)^2 |gr\rangle \langle rr| + \rm{H.c.}\right) + \left((iV_0 + \Gamma)\Omega^2(\Omega+\omega)|rg\rangle\langle rr|+\rm{H.c.}\right)\right],
\end{eqnarray*}
\end{widetext}
and that of subsystems are, 
	\begin{widetext}
\begin{eqnarray*}
\rho_A &=& \frac{1}{\kappa}\left[\left(V_0^2 \left(\Gamma^2+ 2\omega^2 + 4\omega\Omega+ 3\Omega^2\right)+ \left(\Gamma^2 +\Omega^2\right)\left(\Gamma^2+ 2(\Omega+\omega)^2\right)\right)|g\rangle\langle g|\right.\\&&\left. +\Omega^2\left(V_0^2+\Gamma^2+ 2(\Omega+\omega)^2\right)|r\rangle\langle r| + \left(\Omega\left(iV_0^2\Gamma- 2V_0 (\Omega+\omega)^2+ i\Gamma\left(\Gamma^2+ 2(\Omega+\omega)^2\right)\right)|g\rangle\langle r|+\rm{H.c.}\right) \right]\\
\rho_B &=& \frac{1}{\kappa}\left[\left(V_0^2 \left(\Gamma^2+ \omega^2 + 2\omega\Omega+ 3\Omega^2\right)+ \left(\Gamma^2 +2\Omega^2\right)\left(\Gamma^2+ (\Omega+\omega)^2\right)\right)|g\rangle\langle g|\right.\\&&\left. +(\Omega+\omega)^2\left(V_0^2+\Gamma^2+ 2\Omega^2\right)|r\rangle\langle r| - \left((\Omega+\omega)\left(V_0-i\Gamma\right)\left(\Gamma^2+ 2\Omega^2-iV_0\Gamma\right)|g\rangle\langle r|+\rm{H.c.}\right) \right]\\
\end{eqnarray*}
\end{widetext}
where
\begin{equation}
    \kappa = V_0^2\left(\Gamma^2+ 2\omega^2+ 4\Omega\omega+ 4\Omega^2\right)+ (\Gamma^2 + 2\Omega^2)\left(\Gamma^2 + 2(\Omega+\omega)^2\right).
\end{equation}
The purity of the system and subsystems are obtained as:
\begin{widetext}
\begin{eqnarray*}
\rm{Tr}\left(\rho_{AB}^2\right)&=& \frac{1}{\kappa^2}\left[\left(\Gamma^4 + 4\Gamma^2\Omega^2+ 2\Omega^4\right)\left(\Gamma^4 + 4\Gamma^2 (\Omega+\omega)^2+ 2(\Omega+\omega)^4\right)\right.\\&&\left.+ V_0^4\left(\Gamma^4 + 4\Gamma^2\left(2\Omega^2 + 2\Omega\omega+ \omega^2\right) + 2\left(2\Omega^2 + 2\Omega\omega+ \omega^2\right)^2\right)\right. \\&& \left. + V_0^2 \left(\Gamma^6 + 4\Gamma^4\left(2\Omega^2 + 2\Omega\omega+ \omega^2\right)+ 4\Omega^2(\Omega+\omega)^2\left(2\Omega^2 + 2\Omega\omega+ \omega^2\right)\right.\right.\\&&\left.\left.+ 2\Gamma^2\left(\omega^4 + 4\omega^3\Omega+ 11\omega^2\Omega^2+ 14\omega\Omega^3+ 7\Omega^4\right)\right)\right]\\
    \rm{Tr}\left(\rho_{A}^2\right)&=& \frac{1}{\kappa^2}\left[\Omega^4 \left(V_0^2 + \Gamma^2 + 2(\Omega+\omega)^2\right)^2 + 2(V_0^2 + \Gamma^2)\Omega^2\left(\left(\Gamma^2+ 2(\Omega+\omega)^2\right)^2+ V_0^2\Gamma^2\right) \right. \\&& \left. + \left(\Gamma^4 +2\Omega^2(\Omega+\omega)^2+ V_0^2\left(\Gamma^2+ 2\omega^2+ 3\Omega^2\right)+ 4\Omega\omega V_0^2+ \Gamma^2\left(2\omega^2 + 3\Omega^2+ 4\Omega\omega\right)\right)^2 \right]\\
    \rm{Tr}\left(\rho_{B}^2\right)&=& \frac{1}{\kappa^2}\left[(\Omega+\omega)^4 \left(V_0^2 + \Gamma^2 + 2\Omega^2\right)^2 + 2(V_0^2 + \Gamma^2)(\Omega+\omega)^2\left(\left(\Gamma^2+ 2\Omega^2\right)^2+ V_0^2\Gamma^2\right) \right. \\&& \left. + \left(\Gamma^4 +2\Omega^2(\Omega+\omega)^2+ V_0^2\left(\Gamma^2+ \omega^2+ 3\Omega^2\right)+ 2\Omega\omega V_0^2+ \Gamma^2\left(\omega^2 + 3\Omega^2+ 2\Omega\omega\right)\right)^2 \right].
\end{eqnarray*}
\end{widetext}
For $\omega=0$, the expressions for purity become:
\begin{widetext}
\begin{eqnarray}
    \rm{Tr}\left(\rho_{A}^2\right) = \rm{Tr}\left(\rho_{B}^2\right) &=& \frac{1}{\kappa^2}\left[V_0^4\left(\Gamma^4 + 8\Gamma^2\Omega^2 + 10\Omega^4\right) + V_0^2 \left(2\Gamma^6 + 16\Gamma^4\Omega^2 + 32\Gamma^2\Omega^4+ 24\Omega^6\right)\right.\\&&\left. + \Gamma^8+ 8\Gamma^6\Omega^2+ 22\Gamma^4\Omega^4+ 24\Gamma^2\Omega^6+ 8\Omega^8 \right]\\
    \rm{Tr}\left(\rho_{AB}^2\right)&=& \frac{1}{\kappa^2}\left[\left(\Gamma^4+ 4\Gamma^2\Omega^2+ 2\Omega^4\right)^2+ V_0^4\left(\Gamma^4+ 8\Gamma^2\Omega^2+ 8\Omega^4\right)\right.\\&&\left. +2V_0^2\left(\Gamma^6 + 8\Gamma^4\Omega^2 + 14\Gamma^2\Omega^4+ 8\Omega^6\right) \right],
\end{eqnarray}
\end{widetext}
and for $\omega=0$, the parameter $\kappa$ reduces to, 
\begin{equation}
\kappa = V_0^2\left(\Gamma^2 + 4\Omega^2\right)+ \left(\Gamma^2 + 2\Omega^2\right)^2 
\end{equation}
\bibliographystyle{apsrev4-1}
\bibliography{lib.bib}
\end{document}